\documentclass[twocolumn,citeautoscript,pra,superscriptaddress,amsmath,amssymb,8pt]{revtex4-1}

\usepackage{graphicx}
\usepackage{color}
\usepackage[dvipsnames]{xcolor}
\usepackage{upgreek}
\usepackage{float}
\usepackage{upgreek}
\usepackage{amsmath,amssymb}

\newcommand{\degC}[0]{\ensuremath{^{\circ}}C}

\newcommand{\expp}[1]{\exp\left( #1 \right)}

\begin{document}

\title{Spatial mapping of band bending in semiconductor devices using in-situ quantum sensors}

\author{D. A. Broadway}
\thanks{These authors contributed equally to this work.}
\affiliation{School of Physics, University of Melbourne, Parkville, VIC 3010, Australia}
\affiliation{Centre for Quantum Computation and Communication Technology, School of Physics, University of Melbourne, Parkville, VIC 3010, Australia}

\author{N. Dontschuk}
\thanks{These authors contributed equally to this work.}
\affiliation{School of Physics, University of Melbourne, Parkville, VIC 3010, Australia}
\affiliation{Centre for Quantum Computation and Communication Technology, School of Physics, University of Melbourne, Parkville, VIC 3010, Australia}

\author{A. Tsai}
\affiliation{School of Physics, University of Melbourne, Parkville, VIC 3010, Australia}

\author{S. E. Lillie}
\affiliation{School of Physics, University of Melbourne, Parkville, VIC 3010, Australia}
\affiliation{Centre for Quantum Computation and Communication Technology, School of Physics, University of Melbourne, Parkville, VIC 3010, Australia}

\author{C. T.-K. Lew}
\affiliation{School of Physics, University of Melbourne, Parkville, VIC 3010, Australia}
\affiliation{Centre for Quantum Computation and Communication Technology, School of Physics, University of Melbourne, Parkville, VIC 3010, Australia}

\author{J. C. McCallum}
\affiliation{School of Physics, University of Melbourne, Parkville, VIC 3010, Australia}

\author{B. C. Johnson}
\affiliation{School of Physics, University of Melbourne, Parkville, VIC 3010, Australia}
\affiliation{Centre for Quantum Computation and Communication Technology, School of Physics, University of Melbourne, Parkville, VIC 3010, Australia}

\author{M. W. Doherty}
\affiliation{ Laser Physics Centre, Research School of Physics and Engineering, Australian National University, Canberra, ACT 2601, Australia}

\author{A. Stacey}
\affiliation{Centre for Quantum Computation and Communication Technology, School of Physics, University of Melbourne, Parkville, VIC 3010, Australia}
\affiliation{Melbourne Centre for Nanofabrication, Clayton, VIC 3168, Australia}

\author{L. C. L. Hollenberg}
\email{lloydch@unimelb.edu.au}
\affiliation{School of Physics, University of Melbourne, Parkville, VIC 3010, Australia}
\affiliation{Centre for Quantum Computation and Communication Technology, School of Physics, University of Melbourne, Parkville, VIC 3010, Australia}	

\author{J.-P. Tetienne}
\email{jtetienne@unimelb.edu.au}
\affiliation{School of Physics, University of Melbourne, Parkville, VIC 3010, Australia}

\maketitle

\textbf{Band bending is a central concept in solid-state physics that arises from local variations in charge distribution especially near semiconductor interfaces and surfaces~\cite{Zhang2012,Kotadiya2018,Simon2010}. Its precision measurement is vital in a variety of contexts from the optimisation of field effect transistors~\cite{Stathis2006,Zhang2006,Kaczr2018} to the engineering of qubit devices with enhanced stability and coherence~\cite{Weber2010,Kaviani2014,Usman2016}. Existing methods are surface sensitive and are unable to probe band bending at depth from surface or bulk charges related to crystal defects~\cite{Zhang2012,Kronik1999,Ishii2004,Butler2014}. Here we propose an in-situ method for probing band bending in a semiconductor device by imaging an array of atomic-sized quantum sensing defects to report on the local electric field. We implement the concept using the nitrogen-vacancy centre in diamond~\cite{Doherty2013,Dolde2011}, and map the electric field at different depths under various surface terminations. We then fabricate a two-terminal device based on the conductive two-dimensional hole gas formed at a hydrogen-terminated diamond surface \cite{Strobel2004}, and observe an unexpected spatial modulation of the electric field attributed to a complex interplay between charge injection and photo-ionisation effects. Our method opens the way to three-dimensional mapping of band bending in diamond and other semiconductors hosting suitable quantum sensors, combined with simultaneous imaging of charge transport in complex operating devices~\cite{Tetienne2017}.
}
 
The emergence of semiconductor-based quantum sensing technologies in the last decade has opened new opportunities in a range of disciplines across physics, materials science and biology~\cite{Degen2017}. While most existing applications involve sensors that are external to the target sample to be measured~\cite{Rondin2014,Casola2018}, in-situ quantum sensors can also be an extremely valuable resource to study the sample itself by enabling three-dimensional (3D) mapping~\cite{Iwasaki2017}. For semiconductor materials this is especially advantageous as it allows information to be gained on the interactions between surface and bulk defects, which play an important role in semiconductor electronics and in quantum technologies. Here we propose and demonstrate such an application, where in-situ quantum sensors are used to map the electric field near a semiconductor surface, $\vec{\mathcal{E}}$, which is related to band bending via the standard relation 
\begin{equation}\label{Eq: E field}
\vec{\mathcal{E}}(\vec{r}) = \frac{1}{q}\vec{\nabla}E_V(\vec{r}),
\end{equation}
where $E_V(\vec{r})$ is the energy of electrons at the valence band maximum relative to the Fermi level ($q$ is the electron charge). Specifically, we employ the nitrogen-vacancy (NV) centre in diamond~\cite{Doherty2013}, a well established atomic-sized quantum sensing system that was recently shown to be a sensitive electrometer~\cite{Dolde2011,Dolde2014}, although the concept could be applied to other semiconductor systems such as silicon~\cite{Zhang2018} or silicon carbide~\cite{Falk2014}. NV centres can be positioned with a resolution below 1 nm in one dimension and 20 nm in the other two~\cite{Ohno2012,Lesik2016}, making it an ideal candidate for 3D mapping of built-in electric fields. 

\begin{figure*}[t!]
	\includegraphics[width=0.95\linewidth]{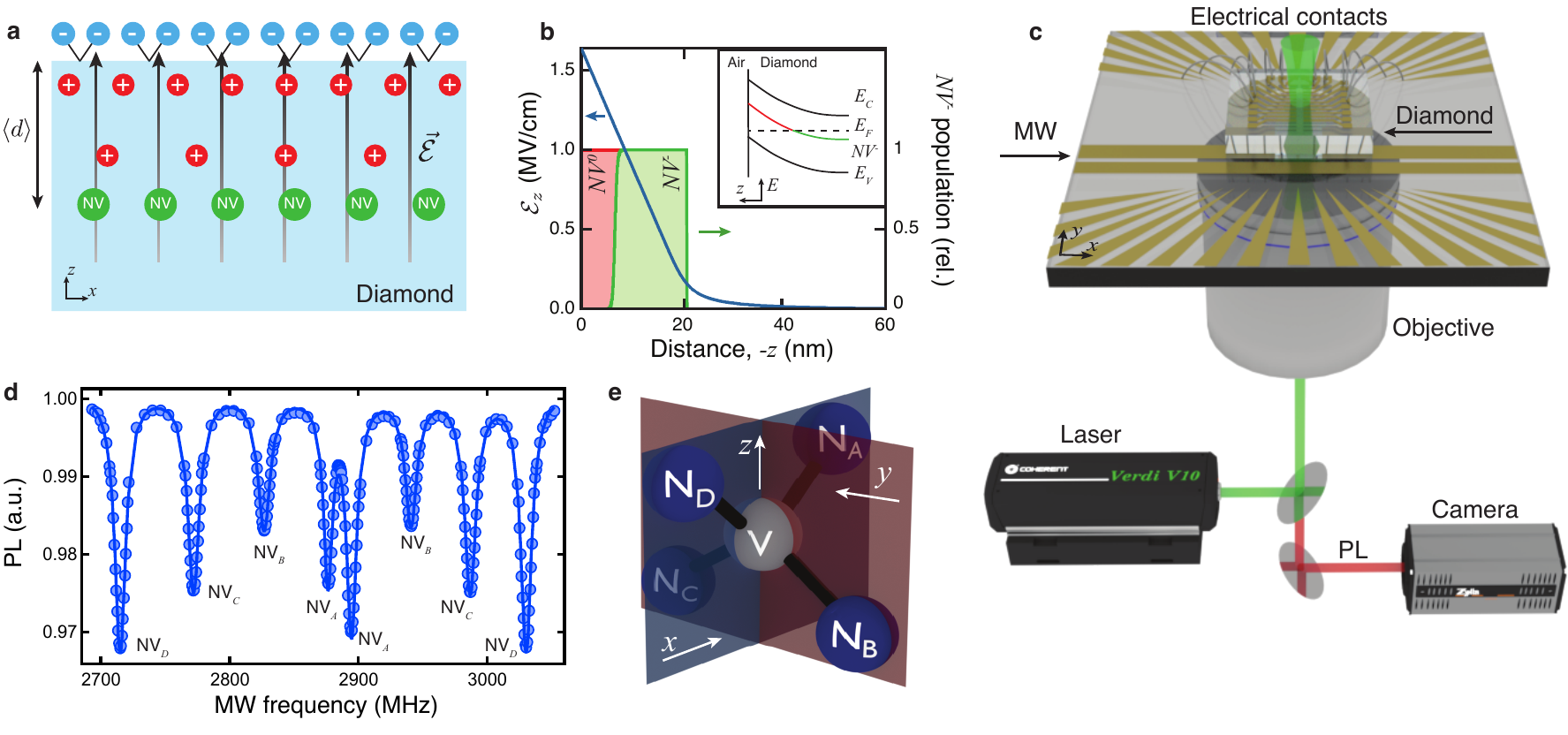}
	\caption{\textbf{Mapping band bending with in-situ quantum sensors.}  \textbf{a}, Principle of the experiment, where NV sensors (green dots) probe the electric field associated with surface band bending, here visualised as a distribution of space charge density. \textbf{b}, Calculated electric field profile for a typical (001)-oriented, oxygen-terminated diamond surface, modelled as a layer of surface acceptor defects with a density of states centred at $1.5$~eV above the valence band maximum~\cite{Stacey2018} and a surface density $D_{sd}=0.1$~nm$^{-2}$ (see details in SI). The implanted substitutional nitrogen and NV defects are taken to be uniformly distributed over the range $d=0-20$~nm (i.e. $\langle d\rangle=10$~nm), with a total areal density of $0.1$~nitrogen/nm$^2$. The green line is the NV$^-$ population at equilibrium, and the green (red) shading represents the region where the NV$^-$ (NV$^0$) charge state is dominant. Inset: corresponding band diagram near the surface, where $E_C$ is the conduction band minimum, $E_V$ the valence band maximum, $E_F$ the Fermi level, and NV$^-$ represents the charge transition level of the NV centre, i.e. NV$^-$ is the stable charge state when this level is below $E_F$. \textbf{c}, Diagram of the experimental set-up showing the diamond sample mounted on a glass slide patterned with gold to allow microwave (MW) injection and interfacing with electrical devices, illumination with a green laser and imaging of the NV red photoluminescence (PL) with a camera. \textbf{d}, Example ODMR spectrum recorded for an ensemble of near-surface NV centres in an oxygen-terminated diamond under a magnetic field $\vec{B}_0$ chosen to be perpendicular (within less than $3^\circ$) to a given NV family (here NV$_A$). Each resonance is labelled according to the corresponding NV orientation, defined in \textbf{e}. The solid line is a multiple-Lorentzian fit. \textbf{e}, The four possible tetrahedral orientations of the NV bond with respect to the sample reference frame $xyz$.} 
	\label{Fig: intro}
\end{figure*}

The principle of the experiment is depicted in Fig.~\ref{Fig: intro}a. At the diamond surface, the bands bend to neutralise any surface charge due to ionised adsorbates or surface defect states, resulting in an electric field perpendicular to the surface with a magnitude $\mathcal{E}_z(z)=\frac{1}{q}\frac{dE_V}{dz}$. To probe this electric field, nitrogen ions were implanted to form NV centres, following a spatial distribution that can be approximated as uniform over the depth range $d=0-2\langle d\rangle$, where $\langle d\rangle$ is the (tunable) mean implantation depth~\cite{Lehtinen2016}. To estimate $\mathcal{E}_z$, we first consider the case of commonly used oxygen-terminated diamond. It was recently found that such samples typically host surface defects that introduce an acceptor level into the band-gap, with densities ($D_{sd}$) as high as 1~nm$^{-2}$~\cite{Stacey2018}. An example of a calculated electric field profile for this scenario (with parameters representative of our implanted samples) is plotted in Fig. \ref{Fig: intro}b, predicting a maximum value at the surface of $\mathcal{E}_z\approx1.6$~MV/cm and a characteristic decay length of $\sim15$~nm. We note that $\mathcal{E}_z$ is positive (i.e., the electric field points towards the surface) which corresponds to the bands bending upward (inset in Fig.~\ref{Fig: intro}b), as expected from a positive space charge density near the surface (see Fig.~\ref{Fig: intro}a). As a consequence, only NVs deeper than a certain threshold (here $\approx7$~nm for $\langle d\rangle=10$~nm) exist in the negatively charged state (NV$^-$) usable for sensing (Fig.~\ref{Fig: intro}b). The expectation value for an electric field measurement, i.e. $\mathcal{E}_z(z)$ averaged over the NV$^-$ distribution, is $\langle \mathcal{E}_z\rangle \approx600$~kV/cm, well in the range of sensitivity of the NV centre~\cite{Dolde2011}. 
 
To measure this electric field, we performed optically detected magnetic resonance (ODMR) spectroscopy on the NV centres, using the experimental set-up depicted in Fig.~\ref{Fig: intro}c. An example ODMR spectrum obtained from a near-surface NV ensemble ($\langle d\rangle\approx10$~nm) is shown in Fig.~\ref{Fig: intro}d. A small magnetic field $\vec{B}_0$ (of magnitude $B_0\sim6$~mT) was applied to split the eight otherwise degenerate electron spin resonances corresponding to the four possible NV defect orientations relative to the diamond crystal (Fig.~\ref{Fig: intro}e), and carefully oriented so as to maximise the sensitivity to electric fields~\cite{Dolde2011} (see SI). The spectrum was fit to extract the eight resonance frequencies, which are then compared to the standard NV spin Hamiltonian including the Zeeman and Stark effects~\cite{Dolde2011,Doherty2012}, allowing us to infer the full vector magnetic and electric fields simultaneously (except for an overall sign ambiguity, i.e. $\vec{\mathcal{E}}$ and $-\vec{\mathcal{E}}$ yield the same ODMR spectrum). Importantly, we found that the measured frequencies could be satisfactorily fit only when accounting for the Stark effect, providing clear evidence of the presence of a non-vanishing electric field (see fit error analysis in SI). For the data shown in Fig. \ref{Fig: intro}d, we obtain $\langle \mathcal{E}_z\rangle=372\pm5$~kV/cm (where we fixed $\mathcal{E}_x=\mathcal{E}_y=0$ to reduce the uncertainty), reasonably close to our estimate. 

\begin{figure}[t!]
	\includegraphics[width=0.95\columnwidth]{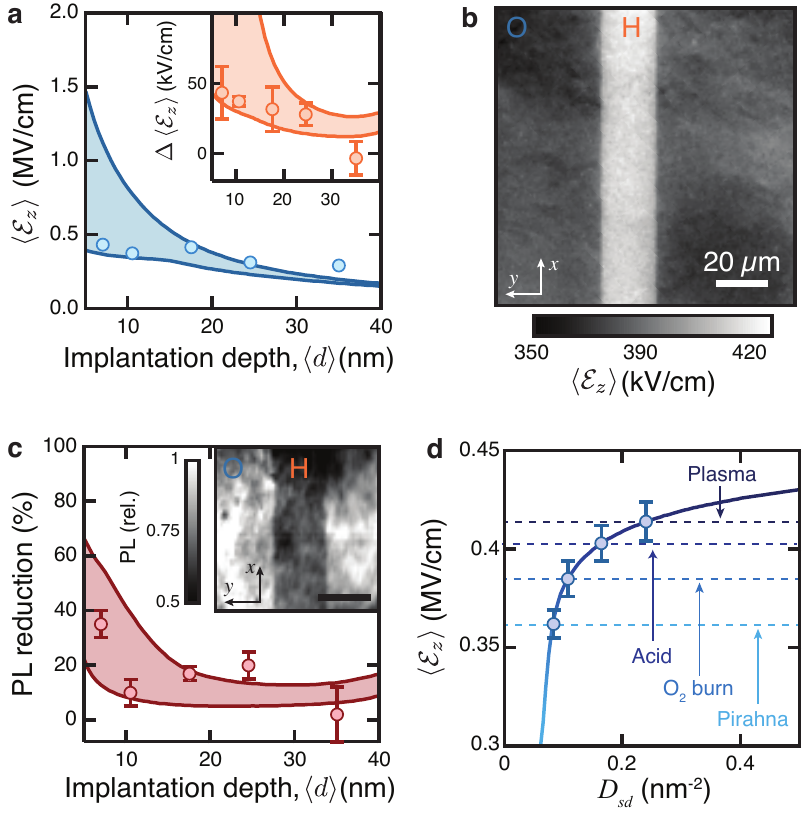}
	\caption{\textbf{Electric field versus implantation depth and surface termination.}  \textbf{a}, Electric field, $\langle \mathcal{E}_z\rangle$, as a function the mean implantation depth, $\langle d\rangle$, for O-terminated diamond. Solid lines: result of the band bending model described in Fig.~\ref{Fig: intro}b with $D_{sd}=0.06$~nm$^{-2}$ (lower curve) and $D_{sd}=1$~nm$^{-2}$ (upper), with the shading representing intermediate values. Inset: Difference $\Delta\langle \mathcal{E}_z\rangle=\langle \mathcal{E}_z\rangle_{\rm H}-\langle \mathcal{E}_z\rangle_{\rm O}$ between the electric field measured for H- and O-terminated diamond. Solid lines: model using a fixed density of charged surface adsorbates, $Q_{sa}=0.07$~nm$^{-2}$, and with $D_{sd}=0.08$~nm$^{-2}$ (upper curve) and $D_{sd}=1$~nm$^{-2}$ (lower). \textbf{b}, $\langle \mathcal{E}_z\rangle$ map of an H-terminated channel on an O-terminated background ($\langle d \rangle\approx10$~nm). \textbf{c}, PL reduction of the H region relative to the O region, as a function of $\langle d\rangle$. Solid lines: model using the same parameters as in inset of \textbf{a}. Inset: PL image of an H-terminated channel ($\langle d \rangle\approx7$~nm). \textbf{d}, $\langle \mathcal{E}_z\rangle$ vs $D_{sd}$ calculated for $\langle d \rangle=17$~nm. The dashed lines and data points indicate the measured $\langle \mathcal{E}_z\rangle$ values for a comparable sample following various surface treatments, performed in order: oxygen plasma (as used to form the O termination in \textbf{a-c}), acid cleaning, oxygen burning and piranha treatment (see details in SI). Vertical error bars: $\pm\sigma$ where $\sigma$ is the standard deviation.}
	\label{Fig: Band bending}
\end{figure} 

To illustrate the 3D mapping capability, we formed NV centres at different depths in distinct diamonds and measured the electric field as explained above. We found that $\langle \mathcal{E}_z\rangle$ decreased from $432\pm10$~kV/cm at $\langle d \rangle\approx7$~nm to $291\pm5$~kV/cm at $\langle d\rangle\approx35$~nm (Fig.~\ref{Fig: Band bending}a), in good agreement with our modelling for $D_{sd}$ in the range $0.06-1$~nm$^{-2}$. We next studied the effect of surface termination on the electric field by forming a hydrogen-terminated (H) channel on an otherwise oxygen-terminated (O) diamond. An example of the resulting $\langle \mathcal{E}_z\rangle$ map is shown in Fig. \ref{Fig: Band bending}b, revealing an increase from $372\pm5$~kV/cm in the O region to $410\pm5$~kV/cm in the H region. This is expected because H-terminated diamond is known to have a lower electron affinity, leading to efficient charge transfer from the diamond material onto acceptor species adsorbed on the surface in ambient air~\cite{Strobel2004}. This leads to increased band bending and hence the increased electric field, beyond the threshold required to form a conductive two-dimensional hole gas (2DHG) near the surface~\cite{Strobel2004,Pakes2014}, which is imaged in an electrical device described below. By fitting our model to the measured increase in $\langle \mathcal{E}_z\rangle$ caused by the H termination (inset in Fig. \ref{Fig: Band bending}a), one can infer the density of charged surface adsorbates (acceptors), $Q_{sa}\approx0.07$~nm$^{-2}$, in good agreement with the value derived from surface resistivity measurements (see SI).

A consequence of the increased band bending is a decrease in the number of NV$^-$ centres, hence a decrease in detected photoluminescence (PL), since the NVs closest to the surface become charge neutral~\cite{Hauf2011}. This is illustrated in Fig.~\ref{Fig: Band bending}c, which shows the PL reduction as a function of $\langle d\rangle$, with an example PL image of a H channel shown in inset. This motivates the need to minimise band bending via surface engineering for quantum sensing applications, where the NV$^-$ to surface distance is critical~\cite{Rondin2014}. As a step towards this goal, we applied various surface treatments in an attempt to reduce the density of surface defects. Starting from a diamond initially O-terminated with an oxygen plasma as previously ($\langle \mathcal{E}_z\rangle\approx414\pm10$~kV/cm), we were able to reduce the electric field to $\approx362\pm6$~kV/cm through a combination of wet and dry treatments, which corresponds to a reduction of $D_{sd}$ by nearly a factor 3 according to our theory (Fig.~\ref{Fig: Band bending}d), and a reduction in the mean NV$^-$ depth from $\approx23$~nm to $\approx19$~nm. These trends are broadly consistent with direct measurements of $D_{sd}$ reported recently~\cite{Stacey2018}. We note that another avenue to reduce $D_{sd}$ is by etching the diamond, as shown in the SI. These results illustrate the value of in-situ electric field measurements, which provide new insights into semiconductor surfaces. 

\begin{figure*}[t!]
	\includegraphics[width=0.9\linewidth]{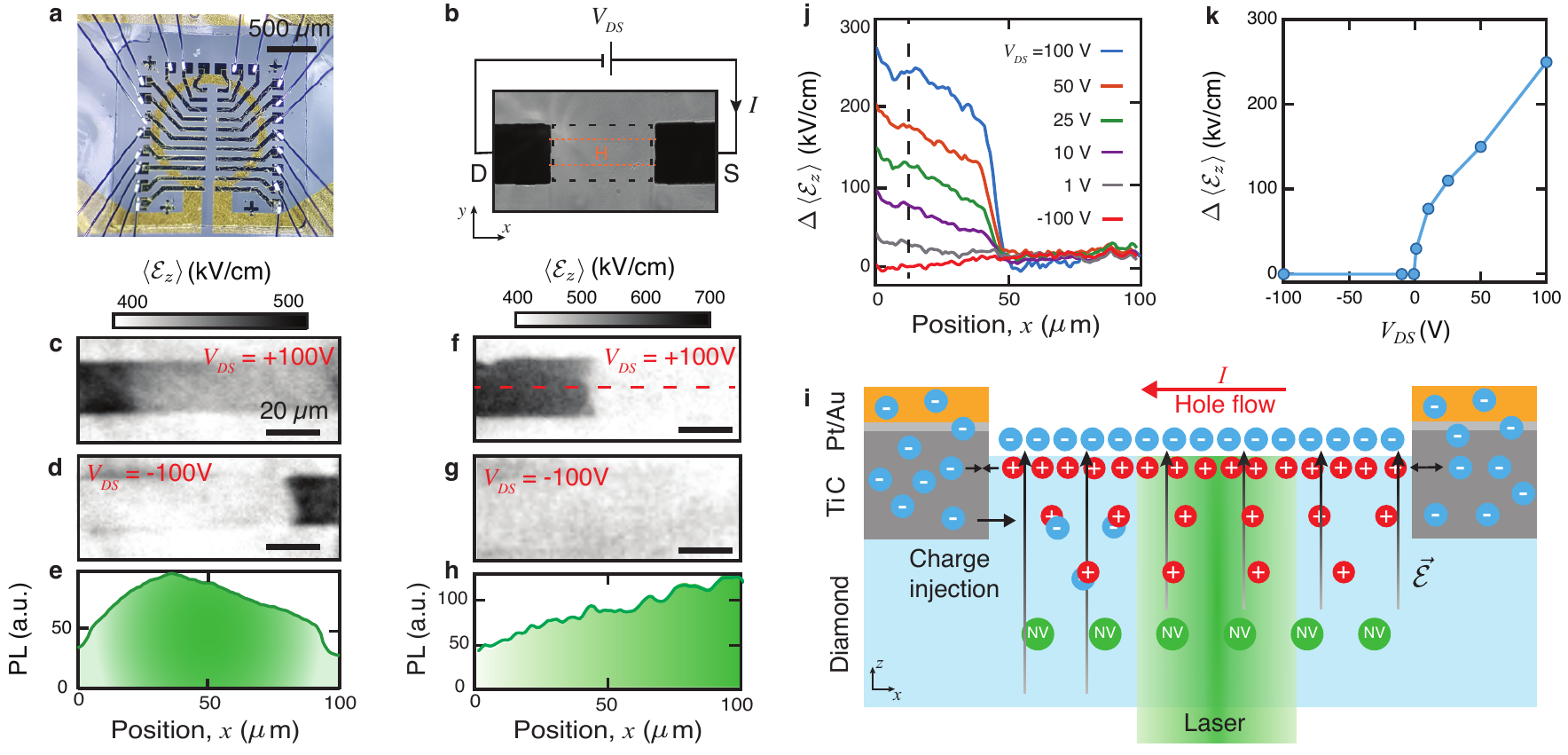}
	\caption{\textbf{Electric field in a two-terminal device.} \textbf{a}, Photograph of the diamond with the fabricated devices. \textbf{b}, Optical bright-field image of a device showing the TiC/Pt/Au contacts in dark, defined as source (S) and drain (D). The optically-transparent 2DHG channel is indicated by orange dotted lines. The black dashed box defines the area imaged in \textbf{c-h}. \textbf{c,d} $\langle \mathcal{E}_z\rangle$ map under an applied voltage $V_{DS}=+100$~V (\textbf{c}) and $V_{DS}=-100$~V (\textbf{d}), corresponding to a current through the channel of $I\approx\pm400~\mu$A. \textbf{e}, Line cut of the PL along the channel, indicative of the laser spot profile (symbolised by the shading under the curve). \textbf{f-h}, Same as \textbf{c-e} but with the laser spot offset by $50~\mu$m to the right. \textbf{i}, Diagram illustrating the different recombination and trapping processes for electrons under laser illumination and bias voltage. The TiC contact extends $\approx15$~nm deep into the diamond (see SI), with the NV centres implanted at a mean depth $\langle d\rangle\approx17$~nm. \textbf{j}, Line cuts of $\langle \mathcal{E}_z\rangle$ taken along the channel (see dashed line in \textbf{f}) for different voltages $V_{DS}$, with the laser spot centred as in \textbf{f-h}. \textbf{k}, Electric field measured at the position indicated by the black dashed line in \textbf{j}, plotted as a function of $V_{DS}$. In \textbf{j,k}, the zero-voltage value was subtracted to show only the increase caused by the applied voltage, $\Delta\langle \mathcal{E}_z\rangle$.} 
	\label{Fig: active device}
\end{figure*}

We now demonstrate mapping of the electric field in an active electrical device consisting of a driven conductive 2DHG channel formed with an H-terminated diamond surface. Two-terminal devices were fabricated where TiC/Pt/Au contacts (source and drain) are connected by a H-terminated channel 100 $\mu$m in length and 20 $\mu$m in width (Figs.~\ref{Fig: active device}a,b). Unexpectedly, upon applying a voltage $V_{DS}=+100$~V an increase in $\langle \mathcal{E}_z\rangle$ was observed (by up to a factor 2) in a well-defined region of the channel extending over $\approx20~\mu$m from the drain (Fig.\,\ref{Fig: active device}c). Upon inversion of the voltage ($V_{DS}=-100$~V), the feature moved to the other contact (Fig.\,\ref{Fig: active device}d) to remain at the hole drain. In addition, we observed an influence of the position of the laser illumination spot used for the measurements. In Figs.\,\ref{Fig: active device}c,d, the laser spot was centred relative to the device as shown by the PL profile in Fig.\,\ref{Fig: active device}e. When the laser was moved by $50~\mu$m towards the right-hand contact (Figs.\,\ref{Fig: active device}f-h), the region of increased electric field appeared to extend further away ($\approx40~\mu$m) from the left-hand contact under positive voltage (Fig.\,\ref{Fig: active device}f), and disappeared completely under negative voltage (Fig.\,\ref{Fig: active device}g).   

These observations are qualitatively interpreted as a combination of two competing effects, illustrated in Fig.\,\ref{Fig: active device}i: the injection of electrons from the drain contact into charge traps in the diamond bulk, increasing the electric field seen by the NV centres, and laser-induced ionisation of these charge traps, allowing the electrons to escape via the conduction band and returning the electric field to its zero-voltage value. This interpretation is corroborated by the negative capacitance measured for these devices (see SI). Line cuts taken at different voltages (Fig.\,\ref{Fig: active device}j) reveal that the electric field decreases steadily from the contact (with a maximum value that increases monotonically with voltage, see Fig.\,\ref{Fig: active device}k) before dropping off abruptly at a specific position independent of voltage (but dependent on laser position), even though the laser intensity increases approximately linearly with position (Fig.\,\ref{Fig: active device}h). This suggests the existence of a strong non-linearity in the ionisation process as a function of laser intensity, possibly due to a change in the ionisation energy of the charge traps caused by ionisation of a second species of defects~\cite{Dhomkar2018}. 
These experiments illustrate how previously un-observable lateral changes in electric field, resulting from a complex contact/diamond junction, can be directly imaged and correlated with the electrical properties of the device (here a negative capacitance). 

Looking forward, our method could be used to validate models of band bending used in semiconductor electronics, or to facilitate the optimisation of surfaces for  semiconductor-based quantum technologies, including the improvement of the charge stability and quantum coherence of near-surface NV centres in diamond, which remains an outstanding problem to date \cite{DeOliveira2017,Stacey2018}. Another exciting prospect is to combine electric field mapping with other quantum sensing modalities, e.g. current flow mapping \cite{Tetienne2017} or noise spectroscopy \cite{Casola2018,Kolkowitz2015}, which would allow simultaneous monitoring of charge transport and band bending, as in a three-terminal field-effect transistor device based on the native 2DHG formed on diamond or on other 2D electronic systems such as graphene.

\textbf{Author contributions.} NV measurements and analysis were performed by D.A.B and J.-P.T, with inputs from M.W.D. The devices were fabricated by N.D and D.A.B, H-terminated by A.T and A.S, and electrically characterised by C.T-K.L and B.C.J. The band bending model was developed by N.D with inputs from D.A.B, J.-P.T, A.S and L.C.L.H. All authors contributed to interpreting the data and writing the manuscript. 

\textbf{Acknowledgements.} We thank Michael Barson, David Simpson and Liam Hall for useful discussions. We acknowledge support from the Australian Research Council (grants CE110001027, DE170100129, FL130100119, DP170102735). J.-P.T  acknowledges support from the University of Melbourne through an Early Career Researcher Grant. D.A.B, A.T, S.E.L and C.T.-K.L are supported by an Australian Government Research Training Program Scholarship. This work was performed in part at the Melbourne Centre for Nanofabrication (MCN) in the Victorian Node of the Australian National Fabrication Facility (ANFF).

\clearpage

\begin{widetext}

\renewcommand{\theequation}{S\arabic{equation}}
\renewcommand{\thefigure}{S\arabic{figure}}
\renewcommand{\thetable}{S\arabic{table}}
\renewcommand{\bibnumfmt}[1]{[S#1]}
\renewcommand{\citenumfont}[1]{S#1}

\section{Experimental methods}

\subsection{Diamond samples}

The NV-diamond samples used in these experiments were made from 4 mm $\times$ 4 mm $\times$ 50 $\mu$m electronic-grade ([N]~$<1$~ppb) single-crystal diamond plates with \{110\} edges and a (001) top facet, purchased from Delaware Diamond Knives, subsequently laser cut into smaller 2 mm $\times$ 2 mm $\times$ 50 $\mu$m plates and acid cleaned (15 minutes in a boiling mixture of sulphuric acid and sodium nitrate). Each plate was then implanted with $^{15}$N$^+$ ions (InnovIon) at a given energy (ranging from 4 to 20 keV, see Table~\ref{T: diamonds}), a dose of $10^{13}$ ions/cm$^2$, and with a tilt angle of 7$^\circ$. Following implantation, the diamonds were annealed in a vacuum of $\sim10^{-5}$~Torr to form the NV centres, using the following sequence \cite{Tetienne2018}: 6h at 400$^\circ$C, 2h ramp to 800$^\circ$C, 6h at 800$^\circ$C, 2h ramp to 1100$^\circ$C, 2h at 1100$^\circ$C, 2h ramp to room temperature. To remove the graphitic layer formed during the annealing at the elevated temperatures, the samples were acid cleaned (15 minutes in a boiling mixture of sulphuric acid and sodium nitrate) before any further surface treatment or fabrication step.

\begin{table}[htb!]
	\begin{tabular}{c | c | c | c | c}
		Diamond & Implantation energy & $d_{\rm max}$ & Figures \\
		\hline
		\#1 & 6 keV & 21~nm & 1d, 2a, 2b, 2c \\
		\#2 & 10 keV & 35~nm & 2a, 2c, 2d, 3, S5-10 \\
		\#3 & 4 keV & 14~nm & 2a, 2c \& inset \\
		\#4 & 14 keV & 49~nm & 2a, 2c \\
		\#5 & 20 keV & 70~nm & 2a, 2c \\
	\end{tabular}
	\caption{List of diamonds used in this work, indicating the implantation energy, $E_{\rm imp}$ (column 2). Column 3: maximum implantation depth used in the modelling, $d_{\rm max}=3.5E_{\rm imp}$, where it is assumed that the defects (N and NV) are uniformly distributed over the range $d=0-d_{\rm max}$, see justification in Sec. \ref{sec:BBstudy}. In the main text, each sample is referred to in terms of the mean implantation depth, $\langle d\rangle=d_{\rm max}/2$. Column 4: figures in which each sample appears.}\label{T: diamonds}
\end{table}

\subsection{Surface treatments}\label{Sec: E and surf treatment}

To form the H-terminated channels measured in main text Figs. 2b,c and 3, the diamond sample was first subject to a soft hydrogen plasma (7 minutes at 400 W, 10 Torr) optimised to make the diamond surface conductive via hydrogenation while preserving the integrity of the NV centres \cite{Hauf2011}. The channels were then protected by a photoresist mask and the sample subject to a soft oxygen plasma (5 minutes at 50 W, 14 MHz RF with a 0.7 Torr O$_2$/Ar pressure) optimised to render the surface non-conductive via oxygen termination, while minimising the amount of etching of the diamond. No topography step was observed in atomic force microscopy (AFM) indicating an etching by the oxygen plasma below our AFM resolution (i.e. $<0.5$~nm). The resulting surface terminations are referred to as hydrogen-terminated (H) and oxygen-terminated (O) in the main text.

In main text Fig. 2d, the effect of other variants of oxygen termination are investigated. Namely, starting with a diamond (sample \#2) oxygen terminated via an oxygen plasma as described previously (labelled as `Plasma' in main text Fig. 2d), we applied the following steps: (i) acid cleaning (15 minutes in a boiling mixture of sulphuric acid and sodium nitrate); (ii) annealing at $500^\circ$C in oxygen-rich atmosphere similar to the process used in Ref. \cite{Lovchinsky2016}; (iii) cleaning in a piranha solution (mixture of 4 ml of sulphuric acid and 2 ml of hydrogen peroxide heated to 90\degC) for 10 minutes. The resulting surfaces are labelled in main text Fig. 2d as `Acid', `O$_2$ burn' and 'Piranha', respectively.    

\subsection{Device fabrication} \label{sec:fab}

The devices such as the one imaged in main text Fig. 3 were fabricated as follows. A stack of Ti/Pt/Au (thickness 10/10/70\,nm) was evaporated onto the diamond (diamond \#2, see Table~\ref{T: diamonds}) masked by a photoresist pattern. After lift-off of the photoresist leaving Ti/Pt/Au contacts (Fig.\,\ref{Fig: Device fab}a), the sample was annealed at 600$^\circ$C for 20 mins in hydrogen gas (10 Torr). At such temperature, Ti atoms are able to diffuse into the diamond, and conversely carbon atoms are able to diffuse into the Ti layer, thus forming a TiC layer extending about 15 nm into the diamond (Fig.\,\ref{Fig: Device fab}b, and Fig. \ref{Fig: Etching} for the depth measurement). The thick Au layer serves as the primary contact material for electrical interfacing, while the Pt layer is introduced to act as a barrier preventing diffusion of Au and Ti atoms across the Ti/Au interface. The interest of such a process is to form a clean quasi-one-dimensional interface with the H-terminated channels subsequently fabricated (Fig.\,\ref{Fig: Device fab}c). Together with the high expected work function of TiC ($\approx 5$~eV \cite{Kim2015}), this results in more consistent formation of Ohmic contacts than with conventional 2D interfaces of H-terminated diamond with Au \cite{Hauf2014} or Ti \cite{Akhgar2016}, for instance. 
After the TiC/Pt/Au contacts were formed, the H-terminated channels were made via the process outlined above, i.e. hydrogen plasma to H-terminate the bare diamond surface, photo-lithography to protect the channels, and oxygen plasma to O-terminate the unprotected diamond surface. Finally, large Cr/Au (10/70\,nm) contact pads partially overlapped with the TiC/Pt/Au contacts were evaporated onto the device for physical wire bonding. 

\begin{figure}[tb!]
	\includegraphics[width=0.85\textwidth]{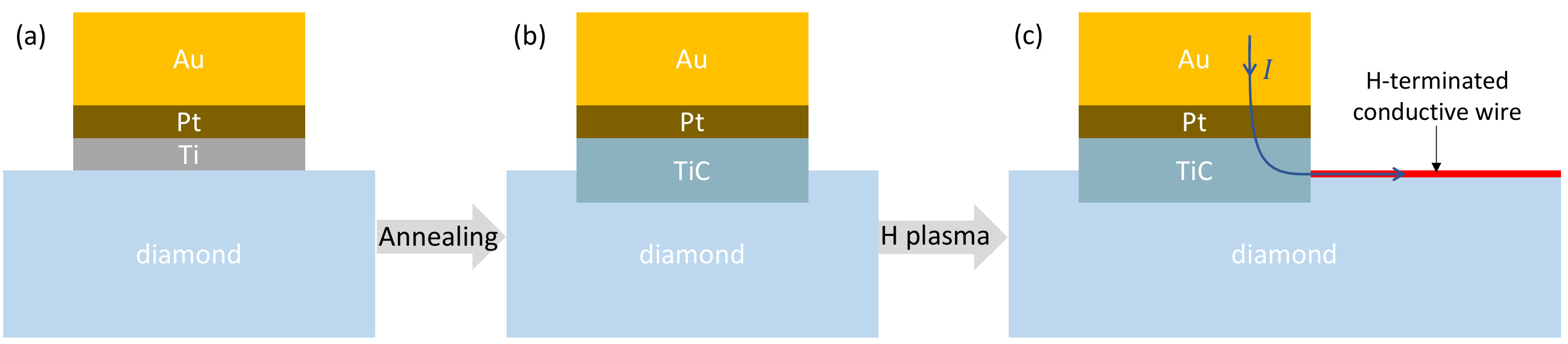}	
	\caption{Schematics illustrating the fabrication process: a Ti/Pt/Au stack is evaporated on the diamond (a) and annealed to form a TiC layer extending into the diamond (b), providing a low-resistivity interface with the conductive 2DHG formed by H-terminating the diamond surface with a hydrogen plasma (c).}
	\label{Fig: Device fab}
\end{figure}

\subsection{Imaging set-up}

The NV imaging set-up is a custom-built wide-field fluorescence microscope similar to that used in Refs. \cite{Simpson2016,Tetienne2017}. The diamonds were glued on a glass cover slip patterned with a microwave waveguide (Fig. \ref{Fig:setup}b), connected to a printed circuit board (PCB, see Fig. \ref{Fig:setup}a) with silver epoxy. The diamond devices were electrically connected to the cover slip via wire bonding, and to the PCB board with silver epoxy. The voltage through the device under study (in main text Fig. 3) was applied using a source-meter unit (Keithley SMU 2450) operated in constant voltage mode. All measurements were performed in an ambient environment at room temperature, under a bias magnetic field $\vec{B}_0$ generated using a permanent magnet (visible in Fig. \ref{Fig:setup}a).

Optical excitation from a 532 nm Verdi laser was gated using an acousto-optic modulator (AA Opto-Electronic MQ180-A0,25-VIS), beam expanded (5x) and focused using a wide-field lens ($f=200$~mm) to the back aperture of an oil immersion objective lens (Nikon CFI S Fluor 40x, NA = 1.3). The photoluminescence (PL) from the NV centres is separated from the excitation light with a dichroic mirror and filtered using a bandpass filter before being imaged using a tube lens ($f=300$~mm) onto a sCMOS camera (Andor Zyla 5.5-W USB3). Microwave excitation was provided by a signal generator (Rohde \& Schwarz SMBV100A) gated using the built-in IQ modulation and amplified (Mini-Circuits ZHL-16W-43+) before being sent to the PCB (see Fig. \ref{Fig:setup}a). A pulse pattern generator (SpinCore PulseBlasterESR-PRO 500 MHz) was used to gate the excitation laser and microwaves and to synchronise the image acquisition. The total CW laser power at the sample was 300 mW, which corresponds to a maximum power density of about 5~kW/cm$^2$ given the $\approx120~\mu$m~$1/e^2$ beam diameter. The optically detected magnetic resonance (ODMR) spectra of the NV layer were obtained by sweeping the microwave frequency while repeating the following sequence: $10~\mu$s laser pulse, $1~\mu$s wait time, $300$~ns microwave pulse (except in Fig.~\ref{Fig: laser} where it was varied); with total acquisition times of several hours typically.   

\begin{figure}[htb!]
	\includegraphics[width=0.7\textwidth]{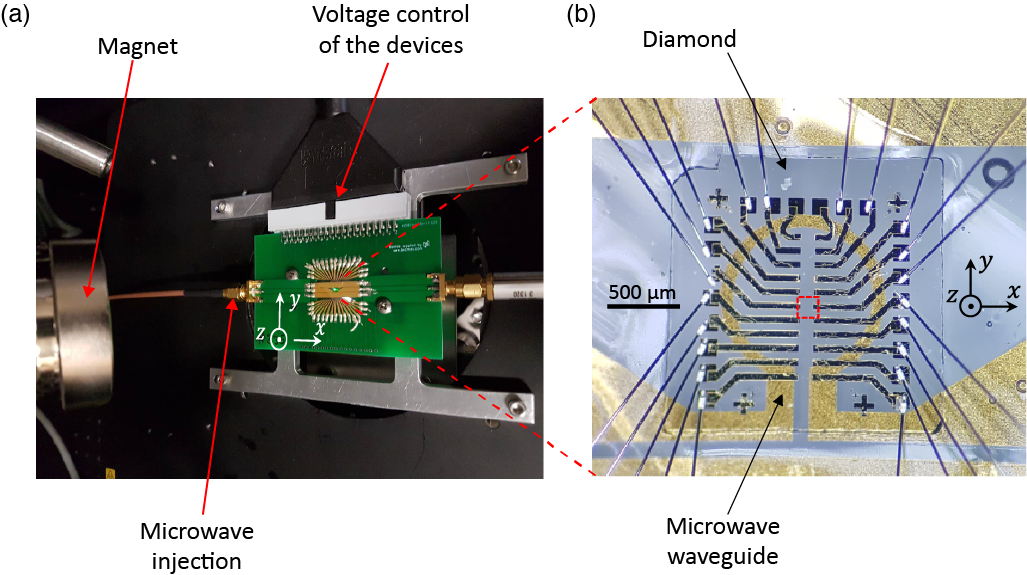}	
	\caption{(a) Photograph showing the PCB with microwave and DC inputs, as well as the permanent magnet used to apply the bias magnetic field $\vec{B}_0$. (b) Photograph of the diamond as mounted. The red box indicates a typical area used for imaging.}
	\label{Fig:setup}
\end{figure}

\section{Electric field mapping with NV ensembles}

In this section, we describe a method to measure the full vector components of the magnetic and electric fields simultaneously, using ensembles of NV centres. To this aim, we first discuss the Hamiltonian of a single NV centre, then show how one can make use of multiple NV centres in a single crystal for maximal sensitivity to electric fields, and finally we describe our fitting and analysis procedure.   

\subsection{Electrometry with a single NV} \label{Sec: Ham}

The spin Hamiltonian of the NV electron spin ($S=1$) in the presence of a magnetic field $\vec{B}=(B_X,B_Y,B_Z)$ and electric field $\vec{\mathcal{E}}=(\mathcal{E}_X,\mathcal{E}_Y,\mathcal{E}_Z)$ is given by 
 \begin{equation}\label{Eq: Ham}
\begin{aligned}
{\cal H} &= (D + k_\parallel \mathcal{E}_Z)(S_Z^2-2/3) + \gamma_{\rm NV}\vec{S}\cdot\vec{B} - k_\perp \mathcal{E}_X (S_X^2 - S_Y^2) + k_\perp \mathcal{E}_Y (S_XS_Y + S_YS_X)
\end{aligned}
\end{equation}
where $\vec{S}=(S_X,S_Y,S_Z)$ are the spin-1 operators, $D\approx2870$\,MHz is the temperature-dependent zero-field splitting, $\gamma_{\rm NV}=28.035(3)$~GHz/T is the isotropic gyromagnetic ratio, and $k_\parallel = 0.35(2)$ Hz cmV$^{-1}$ and $k_\perp= 17(3)$ Hz cmV$^{-1}$ are the electric susceptibility parameters \cite{Doherty2012,Doherty2013}. Here $XYZ$ is the reference frame of the NV defect structure as defined in Ref. \cite{Doherty2014}, where $Z$ is the major symmetry axis defined by the direction joining the nitrogen and the vacancy (which is along a $\langle111\rangle$ crystal direction), and $X$ is a minor symmetry axis defined as being orthogonal to $Z$ and also contained within one of the three reflection planes. 

The frequencies of the $|0\rangle\rightarrow|+1\rangle$ and $|0\rangle\rightarrow|-1\rangle$ spin transitions, $f_\pm$, can be computed by numerically diagonalising the above Hamiltonian. To facilitate the discussion, in what follows we will make use of the approximation 
\begin{equation}\label{Eq: approx}
	f_{\pm} (\vec{\mathcal{E}},\vec{B}) \approx D + k_\parallel \mathcal{E}_Z + 3 \Lambda \pm \sqrt{\mathcal{R}^2 - \Lambda\mathcal{R} \sin\alpha\cos\beta + \Lambda^2} 
\end{equation}
where the terms are defined as
\begin{equation}\label{Eq: approx terms}
\begin{aligned}
\Lambda &= \frac{\gamma_e^2 B_\perp^2}{2D}, \qquad \mathcal{R} = \sqrt{\gamma_{\rm NV}^2 B_Z^2 + k_\perp^2\mathcal{E}_\perp^2}, \qquad B_\perp = \sqrt{B_X^2 + B_Y^2}, \qquad \mathcal{E}_\perp = \sqrt{\mathcal{E}_X^2 + \mathcal{E}_Y^2}, \\
\tan\alpha &= \frac{k_\perp \mathcal{E}_\perp}{\gamma_{\rm NV} B_Z}, \qquad \beta = 2\phi_B + \phi_F, \qquad \tan \phi_B = \frac{B_Y}{B_X}, \qquad  \tan \phi_F = \frac{\mathcal{E}_Y}{\mathcal{E}_X}.
\end{aligned}
\end{equation}
This approximation is valid under the situation where $\Lambda , \mathcal{R} \ll D$ \cite{Doherty2014}, which is a good approximation for the cases explored in this paper.   

As apparent from the expression of $\mathcal{R}$, sensitivity to the transverse electric field ($\mathcal{E}_\perp$) is maximised when the longitudinal component of the magnetic field ($B_Z$) is minimised. To a lesser extent, sensitivity to $\mathcal{E}_\perp$ is also maximised when the transverse magnetic field ($B_\perp$) is minimised, due to the $\Lambda^2$ term in Eq. \ref{Eq: approx}. To illustrate this, a simulation of the shift $\delta f=f_+(\vec{\mathcal{E}})-f_+(0)$ caused by an electric field of strength $|\vec{\mathcal{E}}|=500$~kV/cm is plotted as a function of $(B_Y,B_Z)$ in Fig.\,\ref{Fig: single axis}a. Here the electric field is taken to be along the lab-frame axis $z$ as defined in Fig. \ref{Fig:setup}b, i.e. the [001] direction of the diamond crystal, to mimic the electric field associated with surface band bending. Fig.\,\ref{Fig: single axis}a shows that there is indeed a dramatic decrease in the expected NV frequency shift with $B_Z$, e.g. from $\delta f \approx7$~MHz at zero magnetic field to $\delta f \approx1$~MHz with just $B_Z=10$~G, and a milder effect of the transverse magnetic field, for instance the shift is still $\delta f \approx5$~MHz under $B_Y=50$~G (with $B_Z=0$). There is also a dependence on the orientation of $\vec{B}$ in the transverse ($XY$) plane as shown in Fig.~\ref{Fig: single axis}b, which is related to the direction of both $\vec{B}$ and $\vec{\mathcal{E}}$ relative to the defect minor axis $X$, as captured by the $\cos\beta$ term in Eq. \ref{Eq: approx}. For transverse magnetic field strengths of the order of 50 G, the loss of sensitivity due to a non-optimal angle $\beta$ is relatively mild  (a factor 2 at most), and as such careful alignment of $\vec{B}$ in the transverse plane to match a given direction of $\vec{\mathcal{E}}$ is not critical. We stress that the shift induced by the longitudinal electric field ($\mathcal{E}_Z$) is usually much smaller than that from $\mathcal{E}_\perp$ because $k_\parallel\ll k_\perp$, even though it doesn't decrease with the application of a magnetic field.  

\begin{figure}[t!]
	\includegraphics[width=0.7\textwidth]{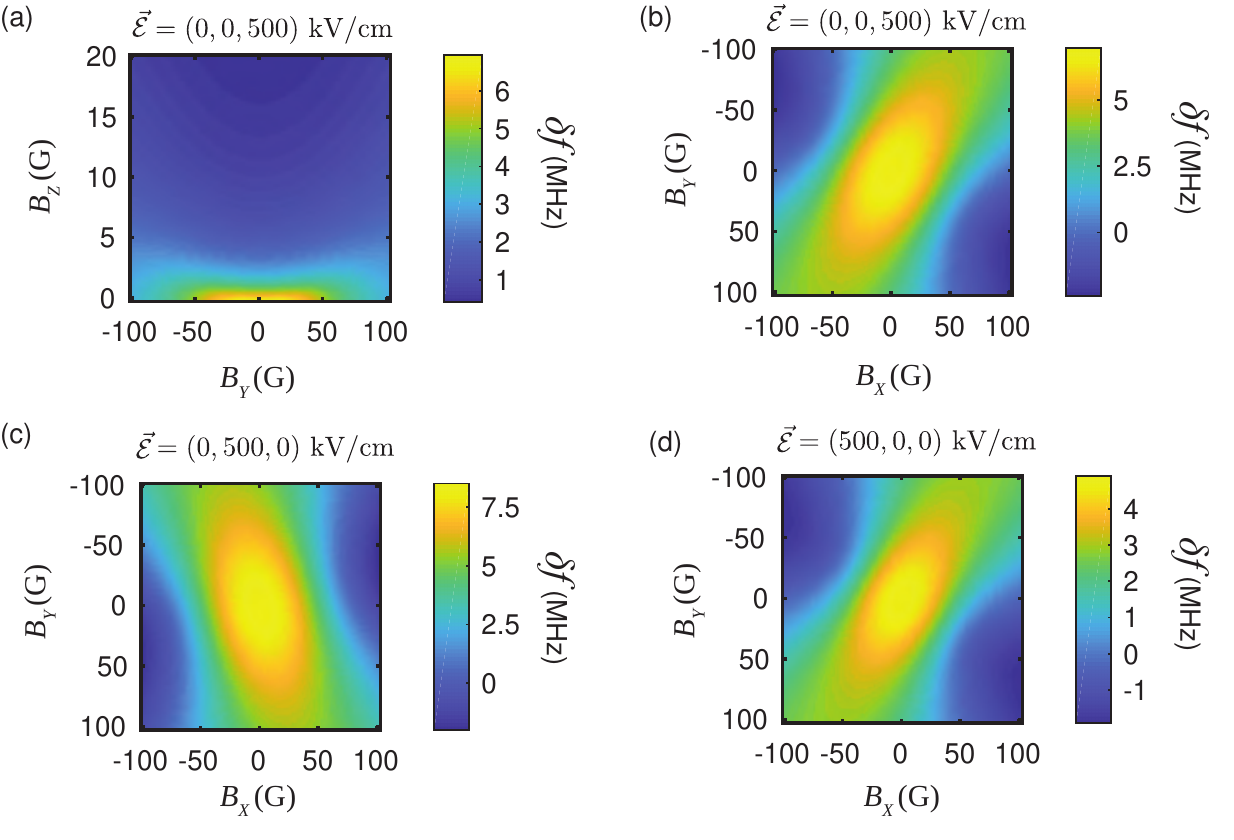}
	\caption{Shift of the $|0\rangle\rightarrow|+1\rangle$ NV spin transition, $\delta f=f_+(\vec{\mathcal{E}})-f_+(0)$, caused by an electric field of strength $|\vec{\mathcal{E}}|=500$~kV/cm, plotted as a function of $(B_Y,B_Z)$ with $B_X=0$ in (a), and as a function of $(B_X,B_Y)$ with $B_Z=0$ in (b-d), where $XYZ$ is the reference frame associated with the NV defect structure, with $Z$ being along a $\langle111\rangle$ direction and $X$ the minor axis \cite{Doherty2014}. $\vec{\mathcal{E}}$ is indicated above each graph in terms of its Cartesian components $(\mathcal{E}_x,\mathcal{E}_y,\mathcal{E}_z)$ in the $xyz$ lab frame as defined in Fig. \ref{Fig:setup}b, i.e. $\vec{\mathcal{E}}$ is aligned along $z$ in (a,b), $y$ in (c) and $x$ in (d).}
	\label{Fig: single axis}
\end{figure}

\subsection{Electrometry with NV ensembles}

Clearly, measuring the two spin transition frequencies $f_{\pm}$ from a single NV centre is not sufficient to infer the six vector components of $\vec{\mathcal{E}}$ and $\vec{B}$ simultaneously, without prior knowledge or additional measurements. On the other hand, using an ensemble of NV centres in a single crystal provides four independent measurements because $Z$ can be along one of the four $\langle111\rangle$ crystal directions, denoted as $\{Z_K\}_{K=A,B,C,D}$. As a result, with eight different frequencies measured there is enough information in principle to infer the six unknown field components, in addition to the zero-field splitting ($D$). Experimentally, this requires all eight frequencies to be spectrally resolvable in the ODMR spectrum, which is achieved by applying a carefully aligned external magnetic field $\vec{B}_0$. 

Precisely, $\vec{B}_0$ is chosen to satisfy several criteria: (i) the direction of $\vec{B}_0$ is chosen perpendicular to one of the $\langle111\rangle$ directions (e.g., $\vec{B}_0 \perp Z_A$) in order to maximise the sensitivity of the corresponding NV centres (family NV$_A$) to electric fields; (ii) the direction of $\vec{B}_0$ within the said transverse plane $X_AY_A$ is varied so that the projections of $\vec{B}_0$ along the four NV axes $\{Z_K\}$ are as distinct to each other as possible; (iii) the amplitude $|\vec{B}_0|$ is chosen as a trade-off between electric field sensitivity which prescribes $|\vec{B}_0|$ to be minimised (see Fig. \ref{Fig: single axis}) and sufficient spacing between adjacent ODMR lines (so they can be resolved given their linewidth). 
To illustrate the last two points, we calculated the minimum splitting between any of the 8 ODMR lines, with frequencies $\{f_i\}_{i=1\dots8}$, defined by
\begin{equation} \label{Eq:separation}
\Delta f_{\min} = \min\left( \sqrt{(f_1 - f_8)^2},  \sqrt{(f_1 - f_7)^2}, \dots , \sqrt{(f_7 - f_8)^2} \right)~,
\end{equation}
as a function of $(B_{X_A},B_{Y_A})$ when $B_{Z_A}=0$ (i.e., $\vec{B}_0 \perp Z_A$) and $\vec{\mathcal{E}}=\vec{0}$, as shown in Fig.\,\ref{Fig: multi axis}a. In our experiments, we aimed for a minimum separation $\Delta f_{\min}\sim20$~MHz nominally, which allows variations in the ODMR frequencies across the field of view of several MHz to be measured (due to variations in the electric or magnetic field generated by the sample or the external magnet). As can be seen in Fig.\,\ref{Fig: multi axis}a, this requires a magnetic field strength of the order of $|\vec{B}|\sim60$~G, with a wide range of directions allowed within the $X_A,Y_A$ plane. A typical experimental ODMR spectrum taken in such conditions is shown in Fig.~\ref{Fig: multi axis}b where $\vec{B}_0=(35,13,-50)$~G, the Cartesian components being expressed in the lab frame $xyz$ defined in Fig. \ref{Fig:setup}b. Fig.~\ref{Fig: multi axis}b also defines our convention to label the 8 NV resonance frequencies and the 4 NV families. With this $\vec{B}_0$, the symmetry axes for the 4 NV families have unit vectors expressed as follows in the $xyz$ lab frame:
\begin{eqnarray}
\vec{u}_{{\rm NV}_A} & = &\left(0,\sqrt{\frac{2}{3}},-\sqrt{\frac{1}{3}}\right) \\
\vec{u}_{{\rm NV}_B} & = &\left(\sqrt{\frac{2}{3}},0,\sqrt{\frac{1}{3}}\right) \\
\vec{u}_{{\rm NV}_C} & = &\left(0,-\sqrt{\frac{2}{3}},-\sqrt{\frac{1}{3}}\right) \\
\vec{u}_{{\rm NV}_D} & = &\left(-\sqrt{\frac{2}{3}},0,\sqrt{\frac{1}{3}}\right)~.
\end{eqnarray}

\begin{figure}[t!]
	\includegraphics[width=0.9\textwidth]{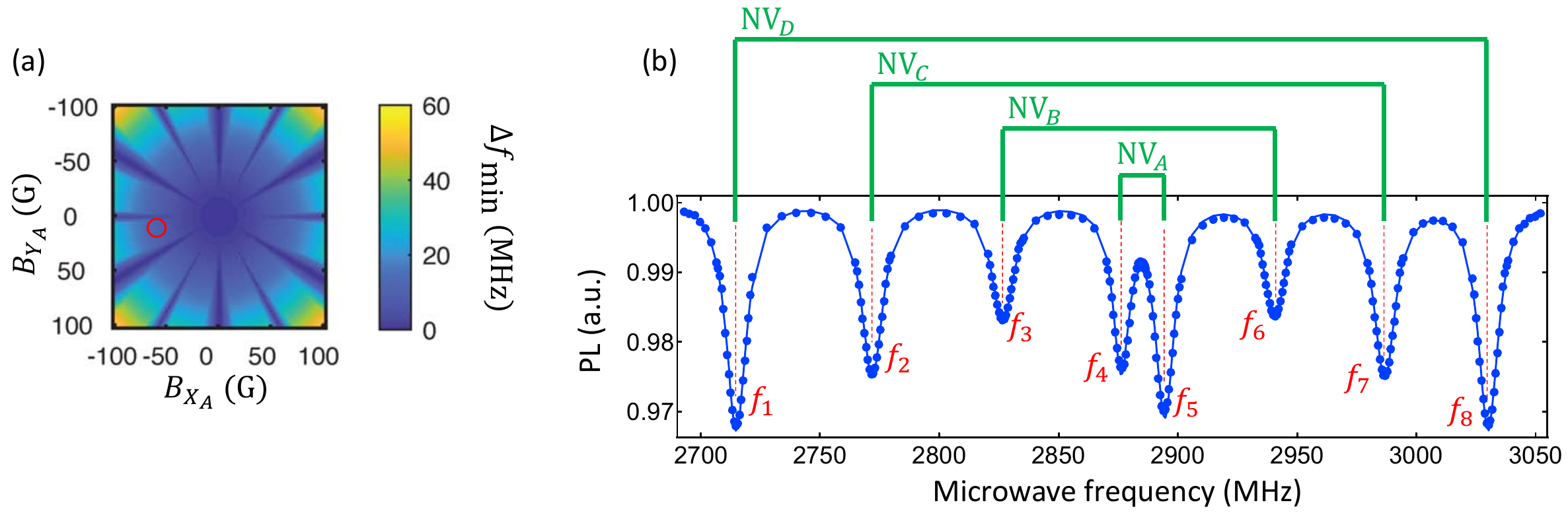}
	\caption{(a) Minimum separation between the 8 ODMR lines as calculated from Eq. \ref{Eq:separation}, as a function of the transverse magnetic field $(B_{X_A},B_{Y_A})$ relative to $Z_A$, which is the symmetry axis of the NV family optimised for electric field sensitivity (i.e., $B_{Z_A}=0$). The red circle indicates the regime used in the experiments, which corresponds to a magnetic field with Cartesian coordinates $\vec{B}_0\approx(35,13,-50)$~G in the lab frame $xyz$. (b) Typical ODMR spectrum measured under this magnetic field $\vec{B}_0$, and conventions for the NV transition frequencies $\{f_i\}_{i=1\dots8}$ and for the NV families $\{{\rm NV}_K\}_{K=A\dots D}$. The solid line is a fit with 8 Lorentzian lines with free amplitudes and widths.}
	\label{Fig: multi axis}
\end{figure}

\begin{figure}[t!]
	\includegraphics[width=\textwidth]{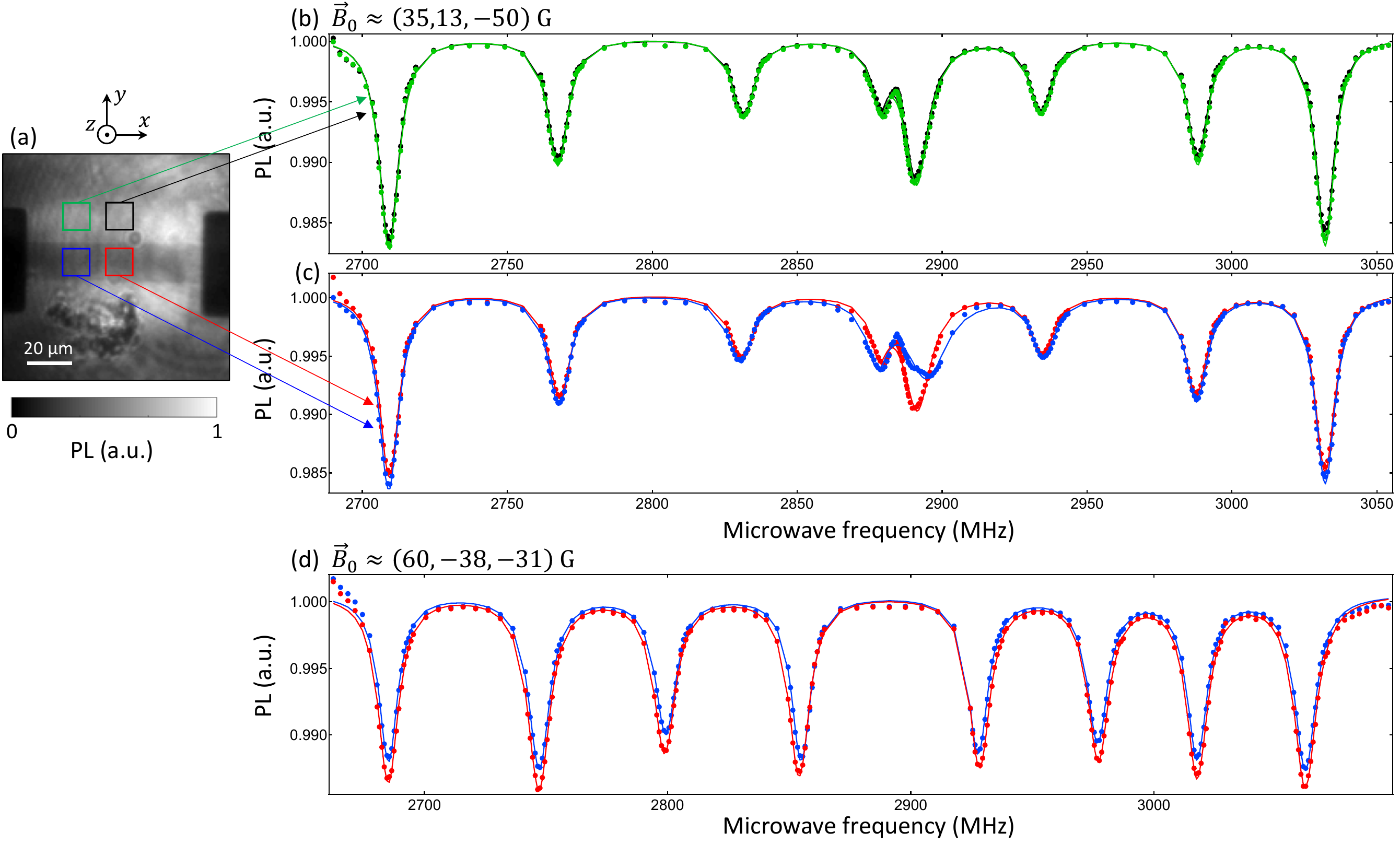}
	\caption{(a) PL image of sample \#2 showing a H-terminated channel between two TiC/Pt/Au contacts. (b,c) ODMR spectra averaged over the regions delimited by the square boxes in (a), with matching colours, recorded under a bias magnetic field $\vec{B}_0\approx(35,13,-50)$~G, a voltage $V_{DS}=+100$~V, and with the laser illumination spot positioned as in main text Fig. 3f. (d) ODMR spectra of the same regions as in (c) but under a magnetic field $\vec{B}_0\approx(60,-38,-31)$~G. In (b-d), the solid lines are data fits with 8 Lorentzian lines with free amplitudes and widths.}
	\label{Fig:Evidence}
\end{figure}

To illustrate how the presence of electric fields affects the ODMR spectrum, we show ODMR data corresponding to the largest electric fields observed in this work, allowing one to readily visualise the induced frequency shifts. We thus consider the situation examined in main text Fig. 3f, where a voltage $V_{DS}=+100$~V was applied to the H-terminated device resulting in an electric-field feature near the drain contact. The PL image of the device is shown in Fig. \ref{Fig:Evidence}a, and ODMR spectra of selected regions are shown in Figs. \ref{Fig:Evidence}b-d under $\vec{B}_0\approx(35,13,-50)$~G (b,c), which is the field optimised for simultaneous vector electrometry and magnetometry and used throughout the paper, and under $\vec{B}_0\approx(60,-38,-31)$~G for comparison (d). The two spectra in Fig. \ref{Fig:Evidence}b are taken outside the H-terminated channel and show no visible shift in the NV frequencies (relative to each other), illustrating the good uniformity of $\vec{B}_0$ over these length scales ($20~\mu$m separate the two plotted regions). Under the conductive channel, however, there is a clear difference between the two spectra (Fig. \ref{Fig:Evidence}c), where the blue spectrum corresponds to the large electric field seen in main text Fig. 3f. The central lines ($f_4$ and $f_5$, family NV$_A$) exhibit shifts by up to 4 MHz (for $f_5$, based on a Lorentzian fit of the whole spectrum) as well as an extra broadening, compared with the reference red spectrum. The other lines are also slightly shifted, which is clearly observable for $f_3$ and $f_6$ (family NV$_B$), but more subtle for the other four lines (NV$_C$ and NV$_D$). Under a non-optimised $\vec{B}_0$ (Fig. \ref{Fig:Evidence}d) where all the NV families have a significant $B_{Z_K}$ component, the shifts are smaller, although they are still visible for NV$_A$. These observations are consistent with a change in electric field between the two regions, which shifts the NV frequencies to an extent that depends on the angle formed between $\vec{B}_0$ and the NV symmetry axis (see Fig.~\ref{Fig: single axis}a). 

\begin{figure}[t!]
	\includegraphics[ width=\textwidth ]{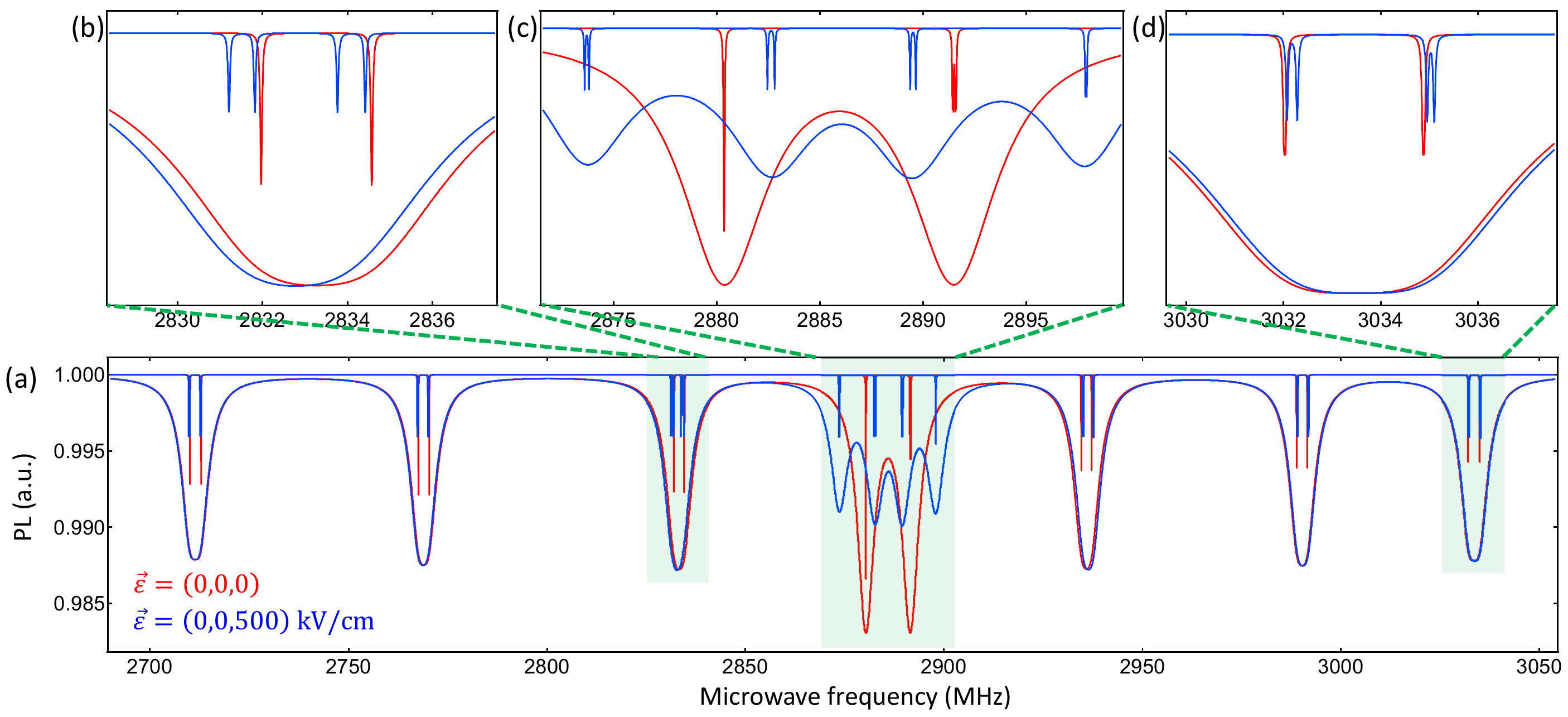}
	\caption{(a) ODMR spectrum obtained by computing the 32 transition frequencies (see details in text) and applying a Lorentzian lineshape with a fixed arbitrary amplitude and a full width at half maximum (FWHM) of 5 MHz (broad lines, comparable to those in the experiment) or 50 KHz (narrow lines, for ease of visualisation). The magnetic field is taken to be $\vec{B}_0\approx(35,13,-50)$~G as in the experiment, while the electric field is null for the red spectrum, and $\vec{\mathcal{E}}=(0,0,500)$~kV/cm for the blue spectrum. (b-d) Close-up views of (a).}
	\label{Fig:ODMR_sims}
\end{figure}

The broadening of $f_4$ and $f_5$ seen in Fig. \ref{Fig:Evidence}c is attributed to a combination of two effects. First, the electric field associated with surface band bending is expected to be non-uniform across the thickness of the NV layer (see Fig.~\ref{BAND_BEND_EX}b), causing an inhomogeneous broadening. The second source of apparent broadening is caused by a splitting of each NV line into two lines corresponding to the two sub-groups of NV centres distinguished by their orientation, i.e. N-V (where the vacancy is closer to the diamond surface than the nitrogen atom) or V-N (the vacancy is closer). In the absence of electric field, the NV transition frequencies are invariant under this inversion, but this is no longer the case in the presence of an electric field \cite{Doherty2014,Michl2014}, especially when the axial component of the magnetic field is vanishing ($B_{Z_A}\sim0$). To illustrate this, we computed the transition frequencies for the 8 possible NV orientations (4 directions + NV/VN inversion) and generated an ODMR spectrum by applying a Lorentzian lineshape to each resonance. For completeness we also included the hyperfine interaction of the NV electron spin with the $^{15}$N nuclear spin (spin-$\frac{1}{2}$), so that the total Hamiltonian of the system is 
\begin{eqnarray} \label{eq:Ham_hyp}
{\cal H} &= (D + k_\parallel \mathcal{E}_Z)(S_Z^2-2/3) + \gamma_{\rm NV}\vec{S}\cdot\vec{B} - k_\perp \mathcal{E}_X (S_X^2 - S_Y^2) + k_\perp \mathcal{E}_Y (S_XS_Y + S_YS_X) \\
&\qquad\qquad - \gamma_n B_Z I_Z + A_{\parallel}S_Z I_Z + A_{\perp} \left( S_X I_X + S_Y I_Y  \right), \notag 
\end{eqnarray}
where ${\bf I}=(I_X,I_Y,I_Z)$ is the nuclear spin operator, $\gamma_n=-4.316$~MHz/T is the nuclear gyromagnetic ratio, and $A_\parallel = 3.03$ MHz and $A_\perp = 3.65$ MHz are the hyperfine parameters  \cite{Felton2009}. One thus has a total of 32 lines in the ODMR spectrum (2 electron spin transitions for each of the 8 possible NV orientations and the 2 possible nuclear spin states), although some of them are nearly degenerate resulting in only 8 lines being usually resolvable under our experimental conditions (due to the intrinsic linewidth of 1-2~ MHz in our samples \cite{Tetienne2018} and additional power broadening). Illustrative simulated spectra are shown in Fig. \ref{Fig:ODMR_sims}a, obtained using $\vec{B}_0\approx(35,13,-50)$~G as in the experiment, and an electric field either null (red spectrum) or of $\vec{\mathcal{E}}=(0,0,500)$~kV/cm (blue). As expected, the presence of an electric field affects especially the central lines (called $f_4$ and $f_5$ according to our previous definition), which split further apart from each other, and additionally split into two sub-lines corresponding to the two possible orientations within family NV$_A$ (i.e. N-V vs. V-N, see Ref. \cite{Doherty2014,Michl2014}), separated by nearly 9 MHz. A higher resolution spectrum (sharp lines in Fig. \ref{Fig:ODMR_sims}) reveal additional splittings by $<400$~kHz caused by the hyperfine interaction, which is highly suppressed for a purely transverse magnetic field. The other lines are also shifted overall by the electric field, where the high-resolution spectrum (Figs. \ref{Fig:ODMR_sims}c,d) reveals a small splitting induced by the orientation inversion (N-V vs. V-N) on top of the usual hyperfine splitting of $\approx3$~MHz.

Looking at the experimental spectrum (blue curve in Fig. \ref{Fig:Evidence}c), we remark the presence of a shoulder on line $f_5$ which may be the signature of this inversion-asymmetry-induced splitting. However, the presence of strong electric field gradients mentioned above acts as a source of broadening which prevents detailed comparison with the theory. Furthermore, while the simulated spectra assumed a constant amplitude for each resonance, this is not the case in reality. Indeed, the transition strength (Rabi frequency) depends on the initial and final states of the NV spin system and on the direction of the microwave field relative to the NV defect structure, and can therefore vary significantly between different transitions and in the presence of an electric field, especially for NV centres with a small axial magnetic field (e.g. NV$_A$). For instance, this explains why the two lines $f_4$ and $f_5$ exhibit different amplitudes even in the low electric field case (see red spectrum in Fig. \ref{Fig:Evidence}c, we verified that $f_4$ had a smaller Rabi frequency than $f_5$, by a factor $\approx2$, under the same microwave power). This complicates the analysis of the experimental spectrum when in the presence of an a priori unknown electric field distribution. Consequently, for imaging purpose we simply fit the experimental spectrum with a sum of 8 Lorentzian lines (solid lines in Figs. \ref{Fig:Evidence}b-d) with free frequencies $\{f_i\}$, amplitudes and widths, thus providing a mean value for each electron spin transition incorporating the effect of hyperfine and inversion-asymmetry splittings. The 8 frequencies are then used to deduce the magnetic and electric field vector components, as will be detailed in the next section.

\subsection{Fitting procedure}

\begin{figure}[t!]
	\includegraphics[ width=\textwidth ]{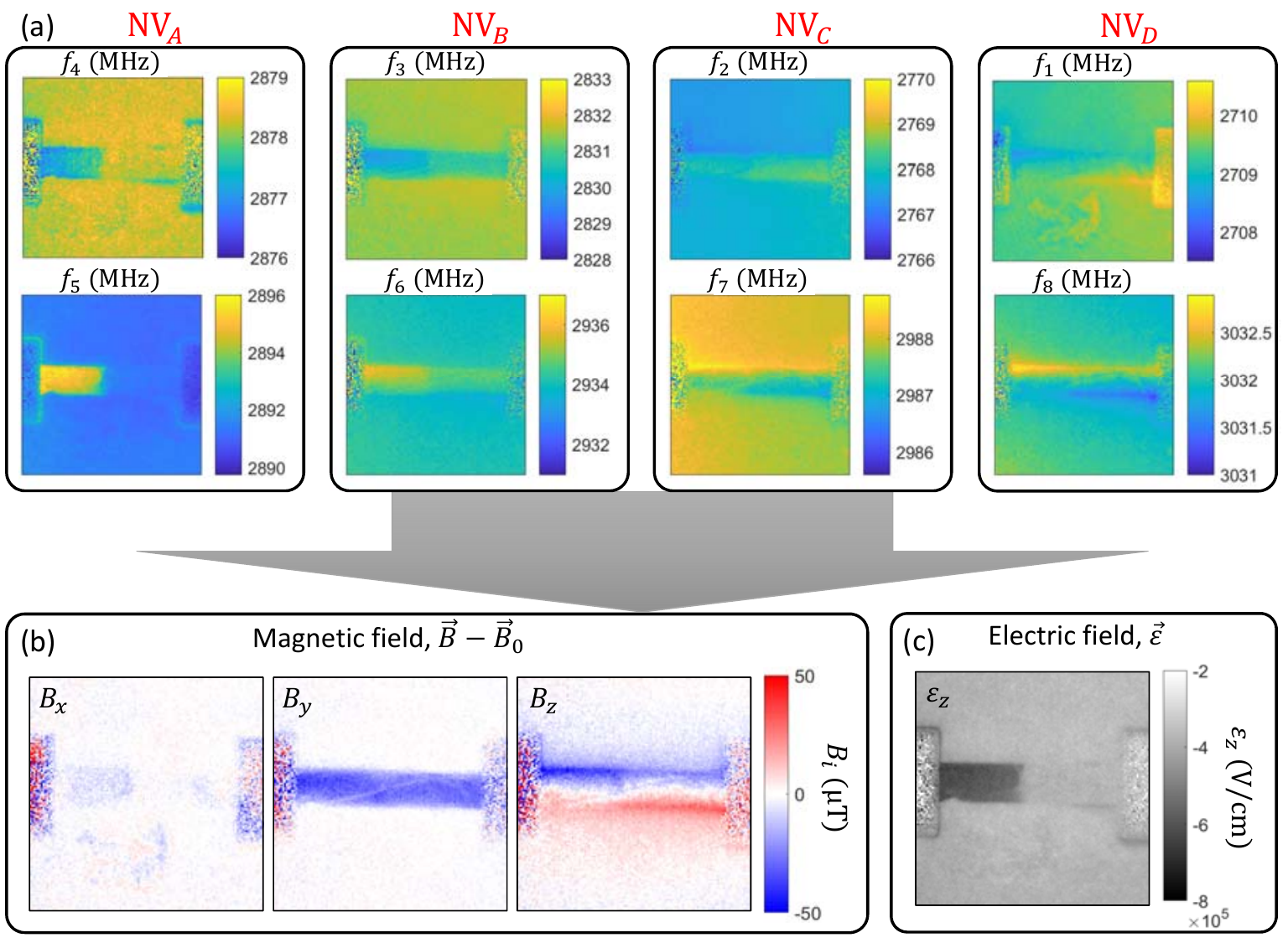}
	\caption{(a) Maps of the frequencies $\{f_i\}_{i=1..8}$ obtained by fitting the ODMR spectrum at each pixel. The region shown is the same as that in Fig. \ref{Fig:Evidence}a, and the bias magnetic field is $\vec{B}_0\approx(35,13,-50)$~G. (b,c) Maps of the magnetic field components (b) and out-of-plane electric field component $\mathcal{E}_z$ (c) obtained after reconstruction. In (b) a plane subtraction was applied to remove the background field $\vec{B}_0$.}
	\label{Fig:Reconstruction}
\end{figure}

To infer the unknown values $(D,B_x,B_y,B_z,\mathcal{E}_x,\mathcal{E}_y,\mathcal{E}_z)$ from the measured frequencies $\{f_i\}_{i=1\dots8}$, we seek to minimise the root-mean-square error function
\begin{equation} \label{Eq:error}
\varepsilon(D,\vec{B},\vec{\mathcal{E}}) = \sqrt{\frac{1}{8}\sum_{i=1}^8 \left[f_i - f_i^{\rm calc}(D,\vec{B},\vec{\mathcal{E}}) \right]^2}
\end{equation}
where $\{f_i^{\rm calc}(D,\vec{B},\vec{\mathcal{E}})\}_{i=1\dots8}$ are the calculated frequencies obtained by averaging over both the NV orientation for the corresponding NV family (i.e., the $Z$ axis is $\pm\vec{u}_{{\rm NV}_K}$), and over the nuclear spin state ($m_I=\pm1/2$), that is,
\begin{eqnarray} 
f_i^{\rm calc}(D,\vec{B},\vec{\mathcal{E}})=&\frac{1}{4}\left[f_i(D,\vec{B},\vec{\mathcal{E}},\vec{u}_{{\rm NV}_K},m_I=+\frac{1}{2})+f_i(D,\vec{B},\vec{\mathcal{E}},-\vec{u}_{{\rm NV}_K},m_I=+\frac{1}{2})\right. \\
& \left. +f_i(D,\vec{B},\vec{\mathcal{E}},\vec{u}_{{\rm NV}_K},m_I=-\frac{1}{2})+f_i(D,\vec{B},\vec{\mathcal{E}},-\vec{u}_{{\rm NV}_K},m_I=-\frac{1}{2})\right]~.
\end{eqnarray}
Each frequency $f_i(D,\vec{B},\vec{\mathcal{E}},\vec{u}_{{\rm NV}_K},m_I)$ is obtained by projecting $\vec{B}$ and $\vec{\mathcal{E}}$ into the NV reference frame knowing the orientation of the $Z$ axis ($\vec{u}_{{\rm NV}_K}$), numerically computing the eigenvalues of the Hamiltonian (\ref{eq:Ham_hyp}), and deducing the electron spin transition frequencies. 

For the fitting (i.e. minimisation of $\varepsilon(D,\vec{B},\vec{\mathcal{E}})$), we used a bound constrained optimisation based on the fminsearch function in Matlab. While it is possible in principle to fit all 7 parameters, we found that the fit would not reliably converge for all the pixels within the field of view, resulting in very noisy electric field images (hence large uncertainty). We conclude that the noise level of our measurements (the uncertainty in the $\{f_i\}$ data is 20-60 kHz for a single pixel, see Table \ref{Table: error each sample}) is insufficient to perform a full vector fit for each pixel. On the other hand, we found that fixing $(\mathcal{E}_x=0,\mathcal{E}_y=0)$ and letting only $\mathcal{E}_z$ free, which is the component expected from surface band bending, solved the convergence issues resulting in much smoother images. In all the data shown in the main text, we therefore fit only the five parameters $(D,B_x,B_y,B_z,\mathcal{E}_z)$, while fixing $(\mathcal{E}_x=0,\mathcal{E}_y=0)$. We also note that because the measurement and model are averaging over the N-V and V-N sub-families, there is an ambiguity in the overall sign of the determined electric field. In other words, $\vec{\mathcal{E}}$ and $-\vec{\mathcal{E}}$ yield the same ODMR spectrum, hence one measures $|\mathcal{E}_z|$ when $(\mathcal{E}_x=0,\mathcal{E}_y=0)$. Solutions to determine the sign of $\vec{\mathcal{E}}$ include using preferentially-oriented NV centres~\cite{Michl2014,Lesik2014} and applying an external electric field~\cite{Dolde2011}.                
As an illustration of the reconstruction process, Fig. \ref{Fig:Reconstruction}a shows the 8 frequency maps obtained for the same area as in Fig. \ref{Fig:Evidence}a. In this example, a voltage $V_{DS}=+100$~V was applied to the device, corresponding to a current $I\approx700~\mu$A flowing through the conductive channel. This current produces a non-trivial magnetic field distribution through the Bio-Savart law, resulting in a complex landscape in the frequency maps. After reconstruction following the procedure outlined above, one obtains the maps of the 3 vector components of $\vec{B}$ (Fig. \ref{Fig:Reconstruction}b) as well as $\mathcal{E}_z$ (Fig. \ref{Fig:Reconstruction}c). The magnetic field maps are consistent with the magnetic field expected from a uniform current density in a flat wire \cite{Tetienne2017}. On the other hand, the $\mathcal{E}_z$ map shows a very different distribution with minimal cross-talk with the magnetic field, illustrating the effectiveness of the reconstruction method.

\subsection{Error analysis}

\begin{table}[b!]
\begin{tabular}{c|c|c|c}
     Sample  & $\varepsilon(D,\vec{B})$ &   $\varepsilon(D,\vec{B},\mathcal{E}_z)$ & $\sigma_{f_i}$ \\
      & (kHz)  &  (kHz)  & (kHz)\\
     \hline
      \#1          & 630  & 45 & 23\\ 
      \#2      & 380  & 60 & 22  \\
      \#3          & 570  & 63  & 47\\ 
      \#4          & 270  & 90 & 53\\ 
      \#5          & 220 & 61 & 20  \\ 
\end{tabular}
\caption{Residual error $\varepsilon$ from fitting the measured ODMR frequencies $\{f_i\}_{i=1..8}$ without including the Stark effect (second column) and when including a normal-to-the-surface electric field $\mathcal{E}_z$ (third column). The fourth column indicates the standard error on the $\{f_i\}$ from fitting the ODMR spectrum with Lorentzian functions. The data shown corresponds to the measurements described in main text Fig. 2a, i.e. O-terminated diamond implanted at different depths, and all values are averaged over a large number of pixels.}
\label{Table: error each sample}
\end{table}

We now discuss the sources of error in our reconstruction method. First, it is interesting to examine the residue $\varepsilon$ after fitting. This is summarised in Table~\ref{Table: error each sample} for the data shown in main text Fig. 2a, which gives the case where the Stark effect is not included in the Hamiltonian, i.e. $\varepsilon(D,\vec{B})$, and the case used in the paper where the Stark effect is included but the electric field is constrained to be perpendicular to the diamond surface, i.e. $\varepsilon(D,\vec{B},\mathcal{E}_z)$. Also indicated in the uncertainty (standard error) in determining the frequencies $\{f_i\}$, denoted as $\sigma_{f_i}$, obtained from the Lorentzian fit to the ODMR spectrum and averaged over the eight lines. Clearly, the data is not well fit without including the Stark effect, with $\varepsilon(D,\vec{B})$ being 5-30 times larger than the measurement uncertainty $\sigma_{f_i}$. Instead, when one includes the Stark effect ($\mathcal{E}_z$ only), one reduces the residue $\varepsilon$ by up to an order of magnitude, which is clear evidence that measurable electric fields are present in our sample. The residue is still a factor 2-3 larger than $\sigma_{f_i}$, which can be due to one or a combination of the following effects. First, the electric field might be not exactly along the $z$ axis, e.g. due to surface roughness, diamond miscut, or non-uniform density of surface or bulk defects, although fitting $\mathcal{E}_x$ and $\mathcal{E}_y$ as well as $\mathcal{E}_z$ did not seem to reduce the residue $\varepsilon$. Second, other corrections to the Hamiltonian may need to be considered, e.g. the effect of mechanical strain in the diamond (see discussion in Sec. \ref{sec:strain}). Third, there may be systematic errors in the measured $\{f_i\}$, for instance due to each ODMR line comprising several sub-lines with uneven amplitudes (they are assumed even in our analysis, see Fig.~\ref{Fig:ODMR_sims} and corresponding discussion) or due to partial overlapping between adjacent lines (leading to systematic errors if the actual line shape is not exactly Lorentzian as it is assumed, which can arise e.g. from the electric field gradient experienced by the NV ensemble). Nevertheless, the fact that simply adding a single parameter ($\mathcal{E}_z$) to the fit function reduces the residue by an order of magnitude suggests that this parameter captures very well the essence of the problem.      

We listed above the possible causes for systematic errors in the measurements. On top of that, there is a statistical error in the measured $\{f_i\}$ that originates from photon count noise in the ODMR spectrum, which is usually close to the photon shot noise limit in our experiments but can also be affected by readout noise and dark counts of the camera depending on the exact conditions. To characterise this noise, we simply calculate the standard deviation $\sigma_{f_i}$ from a large ensemble of pixels, which is found to be in the range 20-60 kHz in our experiments (see Table~\ref{Table: error each sample}). This uncertainty on the $\{f_i\}_{i=1..8}$ translates into an uncertainty on $\mathcal{E}_z$. Again, we characterise this by taking the standard deviation of the $\mathcal{E}_z$ map. Doing so on a small region gives an uncertainty of the order of 1 kV/cm, i.e. less than 1~\% of the mean value, corresponding to the statistical (pixel-to-pixel) noise. By evaluating the standard deviation over larger areas, one captures statistical noise as well as actual spatial variations of $\mathcal{E}_z$, due e.g. to spatial variations in the density of surface defects. Such variations can be seen in main text Fig. 2b and are of the order of 10 kV/cm, i.e. a few percents of the mean value. This is the uncertainty quoted in the main text and shown as error bars in the main text figures.

Finally, we discuss the consequences of the Stark effect from built-in electric fields on the accuracy of magnetometry measurements. In our measurements, because $\vec{B}$ is treated as a fit parameter, neglecting the Stark effect leads to a biased estimation of $\vec{B}$ in order to minimise $\varepsilon(D,\vec{B})$. Taking sample \#1 from Table~\ref{Table: error each sample} as an example, we obtain $(D=2870.2~{\rm MHz},B_x=3672~\mu{\rm T},B_y=1027~\mu{\rm T},B_z=-4987~\mu{\rm T},\mathcal{E}_z=378.4~{\rm kV/cm})$ when including $\mathcal{E}_z$, against $(D=2870.2~{\rm MHz},B_x=3699~\mu{\rm T},B_y=1025~\mu{\rm T},B_z=-4976~\mu{\rm T})$ when neglecting the Stark effect. This shows that neglecting the Stark effect may lead to systematic errors of the order of $30~\mu$T in magnetometry measurements, motivating the need for precise characterisation of built-in electric fields for high-precision magnetometry applications, and minimising the sensitivity to electric fields by careful alignment of an external bias magnetic field.

\subsection{On the effect of strain} \label{sec:strain}

Mechanical strain acts as an effective electric field in the NV spin Hamiltonian \cite{Doherty2012}, which simply adds to the true electric field and therefore may in principle contribute to the electric fields measured in this work. In this section, we estimate the effect of strain in our samples induced by bulk defects, surface defects as well as dilational dislocations, and conclude that it is negligible compared to the electric field due to surface band bending. 

\subsubsection{Strain induced by bulk point defects}

In the simplest picture, a point defect is an isotropic point source of expansion or contraction of the lattice~\cite{Stoneham1968}. In continuum mechanics, the displacement field $\vec{u}$ of this expansion/contraction is analogous to the electric field of a point charge and takes the form~\cite{Stoneham1968}
\begin{equation}
    \vec{u}=\frac{A_0}{\left|\vec{r}-\vec{r}_0\right|^3} \left( \vec{r}-\vec{r}_0 \right),
\end{equation}
where $A_0$ is known as the defect strength and $\vec{r}_0$ is the position of the defect. This displacement field is the solution of the elastic equation
\begin{equation}
    \vec{\nabla }\cdot \vec{u}=4 \pi A_0 \delta  \left(\vec{r}-\vec{r}_0\right).
\end{equation}
Since the displacement field induced by a point defect is irrotational $\vec{\nabla }\times \vec{u}=\vec{0}$, the displacement field can be expressed in terms of a scalar potential, $\Phi$, via $\vec{u}=-\vec{\nabla }\Phi$. In which case, the elasticity equation becomes (see references \cite{Love2011} for background)
\begin{equation}
    -\nabla ^2\Phi =4 \pi A_0 \delta\left(\vec{r}-\vec{r}_0\right).
\end{equation}
Note the parallel between $\Phi$ and the electric potential.

The defect strength is related to the change in volume $\Delta V_0$ induced by the defect \cite{Stoneham1968}
\begin{equation}
    A_0=\frac{\Delta V_0 (\nu +1)}{12 \pi  (1-\nu )}
\end{equation}
where $\nu$ is the Poisson ratio of the solid. The change in volume can be obtained from X-ray crystallography that determines the solid's lattice parameter as a function of the defect concentration or by ab initio calculations.

The components of the strain field are defined with respect to the displacement field via
\begin{equation}
    \epsilon _{ij}=\frac{1}{2}\left(\frac{\partial u_i}{\partial r_j}+\frac{\partial u_j}{\partial r_i}\right)
\end{equation}
where $i,j=x,y,z,u_i$ denotes the $i$-vector component of the displacement field and $r_i$ denotes the $i$-coordinate. The stress is calculated from the strain via
\begin{equation}
    \begin{aligned}
    \sigma_{ii} &=\lambda \text{Tr}\overleftrightarrow{\epsilon} + 2 \mu \epsilon_{ii}\\
    \sigma_{ij} &= \mu \epsilon_{ii} \quad \text{for  } i\neq j
    \end{aligned}
\end{equation}
 where $\lambda$ and $\mu$ are Lame's constant and the shear modulus, respectively.

Assuming that the displacements due to different defects are sufficiently small that they may be added in superposition, the elasticity equation for a continuous number density $\vec{\rho}$ of defects is simply
\begin{equation}
    -\nabla ^2\Phi = 4 \pi A_0 \rho (\vec{r})
\end{equation}

Consider now a semi-infinite slab in the $xy$ plane, extending vertically from $z=-h/2$ to $z=h/2$. If the slab has a uniform density $\rho$ of defects, the solutions of the elasticity equation immediately follow as 
\begin{equation}
    \begin{aligned}
    \Phi &= -2 \pi A_0 \rho z^2 \\
    \vec{\mu } &= 4 \pi A_0 \rho z \hat{z} \\
    \epsilon_{zz} &= 4 \pi A_0 \rho \\
    \sigma_{zz} &= 4 \pi (\lambda +2 \mu ) A_0 \rho   \\
    \sigma_{xx} &= \sigma_{yy} = 4 \pi \lambda A_0 \rho.
    \end{aligned}
\end{equation}

All other strain and stress components are zero. Thus, the defects produce uniform stress and strain that is principally in the vertical direction (assuming $\mu \gg \Lambda$) and proportional to the density of defects.

For the N defect in diamond, $A_0 \rho = 1.67113 \times 10^{-8}$ using the x-ray data $\Delta V_0 \rho = 0.42 \times 10^6$ ppm$^{-1}$~\cite{Lang1991} and the diamond Poisson ratio $\nu=0.2$. Using $\lambda = 85$~GPa and $\mu = 536$~GPa, the stress components as a function of concentration (units: GPa/ppm) are $242.806 \times 10^{-6}$ and $17.838  \times 10^{-6}$ repectively. Since the spin-stress susceptibility parameters are on the order of 1 MHz/GPa, and the N concentrations in our implanted samples are on the order of 10 ppm, we get that the NV frequency shifts induced are $\sim10$~kHz, equivalent to $\sim1$~kV/cm in terms of effective electric field. This is smaller than our measurement uncertainty and therefore negligible. The strengths of other common defects (N$^+$, NV$^-$ and NV$^0$) are smaller than N$^0$, as predicted by the ab initio calculations in \cite{Biktagirov2017}. Thus, even considering multiple defect species, it is unlikely that strain will induce a significant shift in the spin resonances.

\subsubsection{Strain induced by surface defects}

To model surface defects, reconsider the slab but with a defect density that decays exponentially from the surface
\begin{equation}
    \rho(z)=\rho _0\left(e^{-\left|z-\frac{h}{2}\right|/L}+e^{-\left|\frac{h}{2}+z\right|/L}\right)
\end{equation}
In this case, the elasticity solutions are
\begin{equation}
    \begin{aligned}
    \Phi &=-4\pi L^2A_0\rho_0\left(e^{-\left|z-\frac{h}{2}\right|/L}+e^{-\left|\frac{h}{2}+z\right|/L}\right) \\
    \vec{\mu } &= 4\pi LA_0\rho_0\left(e^{-\left|z-\frac{h}{2}\right|/L}+e^{-\left|\frac{h}{2}+z\right|/L}\right)\hat{z} \\
    \epsilon _{zz} &= 4\pi A_0\rho_0 \left(e^{-\left|z-\frac{h}{2}\right|/L} + e^{-\left|\frac{h}{2}+z\right|/L}\right) \\
    \sigma _{zz} &= 4\pi \lambda +2 \mu A_0\rho_0 \left(e^{-\left|z-\frac{h}{2}\right|/L} + e^{-\left|\frac{h}{2}+z\right|/L}\right) \\
    \sigma_{xx} &= \sigma_{yy} = 4\pi \lambda A_0\rho_0\left(e^{-\left|z-\frac{h}{2}\right|/L} + e^{-\left|\frac{h}{2}+z\right|/L}\right)
    \end{aligned}
\end{equation}
Again, the principal stress is in the vertical direction and is proportional to the density of defects. As the density decays exponentially from the surface,  defects densities that are highly localised to the surface do not generate stress inside the solid, and their effect on the NV spin resonances can therefore be neglected.

\subsubsection{Stress induced by dilational dislocations}
A dislocation that induces pure compression/dilation is equivalent to a line of point defects with uniform density. So, if we say that the dislocations have a uniform density in the transverse plane of a slab, this is equivalent to a uniform plane of point defects. Since we have concluded that the stress/strain at a given height in the slab depends only on the local density of defects, and we expect dislocations induced by polishing/surface damage to be localised to the surface, then we can conclude that dislocations do not influence the NVs sufficiently below the surface. Thus, regardless of their strength, dislocations may also be ignored.

\section{Supplementary data}

\subsection{Electrical characterisation of the devices} \label{sec:elec charac}

\begin{figure}[b!]
	\includegraphics{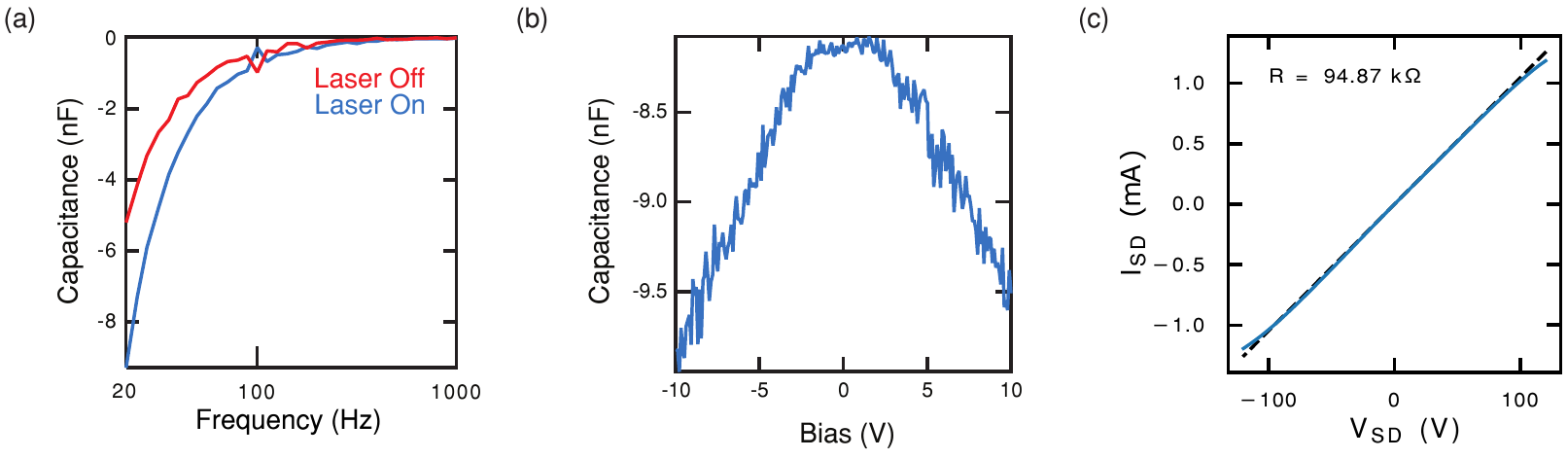}
	\caption{Electrical characterisation of a representative H-terminated channel between two TiC/Pt/Au contacts. (a) Capacitance under a bias voltage $V_{DS}= 10$~V, as a function of the frequency of the AC signal. (b) Capacitance at 20 Hz as a function of the bias voltage (laser on). (c) Source-drain current as a function of the applied DC voltage.}
	\label{Fig: E measurements}
\end{figure}

The electrical properties of the hydrogen terminated channels were characterised first by DC measurements to determine the device resistance (Keithley 2400 SMU), and then with a small AC voltage test signal ($10$\,mV peak to peak) applied by a LCR meter (HP 4284A) to determine the complex device impedance (Z).  The resistance of the devices was modified by the $532$~nm laser illumination required by NV measurements. Upon illumination the resistance of all devices dropped from $R \approx 90$\,k$\Omega$ to $R \approx 75$\,k$\Omega$ ($V_{DS}= 10$~V). Green laser illumination is known to optically pump charge carriers in nitrogen doped diamond through deep defects and surface defects~\cite{Nesladek1998}. The photo-induced carriers in our devices were observed to have a long lifetime with photo-currents taking minutes to decay. Prolonged laser exposure resulted in a gradual and permanent increase in the device resistance. After multiple days of continuous laser illumination the devices typically exhibited dark resistances of $R > 100$\,M$\Omega$ (illuminated resistances of $R \approx 500$\,k$\Omega$), which can be explained by a decrease in dark hole density with no change in the photo-induced carrier density. The reduction of hole density indicates a loss of surface acceptors either by laser induced removal of the hydrogen termination or by the formation of a surface contamination that prohibits transfer doping of the diamond.  

The TiC contacts to the hydrogen terminated ribbon exhibited an Ohmic response within the applied DC bias range of $V_{DS} = 100$ to $V_{DS} = -100$\,V shown in Fig.~\ref{Fig: E measurements}c.  The sub linear response seen above $V_{DS} = 100$\,V is consistent with a drift saturation velocity of $ v_{\rm drift} = 5\times10^6$\,cm\,s$^{-1}$ \cite{Hauf2014}. Despite these stable Ohmic contacts, AC measurements revealed a non-zero parallel capacitance in the diamond device.  At low frequencies the device exhibits a large negative capacitance (Fig.~\ref{Fig: E measurements}a, red line), which becomes positive ($\approx 10$\,pF) above $1$\,kHz.  We interpret this capacitance as a result of the geometrical differences between the TiC contacts, which extend $\Delta z \approx 15$\,nm into the diamond, and the conductive H-terminated channel, which is confined within $z\approx 1$\,nm~\cite{Hauf2014} of the surface, resulting in charge build up at subsurface TiC-diamond interface, below the 2D hole gas.  The frequency response of the negative capacitance is inconsistent with a parasitic inductance, and is understood to result from the transient relaxation current ($\delta j$) after a small voltage step not decreasing monotonically (i.e. $\frac{d\delta j}{dt} > 0$ at some time, $t$, after the voltage step).   The longer times required before $\frac{d\delta j}{dt} > 0$ results in the measured negative capacitance disappearing with high frequency measurements, as the voltage changes are quick enough to ensure $\frac{d\delta j}{dt} < 0$~\cite{Ershov1998, Jonscher1986}. The microscopic cause of this unusual transient response is device specific, for our devices we postulate that the negative capacitance is a result of carrier trapping dynamics across the metal TiC and p-type diamond interface similar to the cause of negative capacitance seen in silicon based p-n junctions~\cite{Lemmi1999}.  

Fig.~\ref{Fig: E measurements}b shows the low frequency negative capacitance response to the applied DC bias. The symmetry around $V_{\rm DS} = 0$ suggests that TiC-diamond interfaces at both contacts behave consistently under a particular bias direction.  Electric field images of the biased interfaces (main text Fig.~3) suggest that this negative capacitance corresponds with the increased electric field and occurs when the interface is forward biased, i.e. when the p-type H-terminated channel is positively biased with respect to the TiC. This indicates that either holes are injected into the TiC, electrons are injected into the diamond or both. As TiC is typically metallic \cite{Leroy2006}, these holes would rapidly recombine, however electrons injected into the conduction band far away from the diamond surface may have a longer lifetime. Trapping of these electrons in ionised donor defects would re-distribute charge within the diamond resulting in a change in the electric field at the NV layer and cause a positive transient response in the current due to the hole density increasing to maintain charge neutrality with the adsorbed acceptor layer. Laser illumination would pump any filled trap states reducing the measured electric field, and increase the magnitude of the positive component of the current transient which results in the increase in the negative capacitance at low frequencies shown in Fig.~\ref{Fig: E measurements}a (blue line). 

\subsection{Electric field vs laser intensity}

To assess the effect of the laser used for the NV measurements on the band bending (through photo-ionisation of bulk defects), we measured the electric field $\langle \mathcal{E}_z\rangle$ as a function of laser intensity for diamond \#2 (with no applied voltage). Precisely, we kept the peak laser power (300 mW) and laser pulse duration ($t_\text{laser}=10~\mu$s) constant and varied the dark time $t_\text{dark}$, which varied the laser duty cycle and hence the average power (Fig.\,\ref{Fig: laser}a) according to
\begin{equation}
\text{Duty cycle} = \frac{t_\text{laser}}{t_\text{dark}+t_\text{laser}}.
\end{equation}  
Since photo-currents have been measured to be long lived (minutes compared to the microsecond time scale of the laser pulsing sequence), we expect varying the average laser power via pulsing to be equivalent to varying the laser power in a CW experiment. We observed a decrease in $\langle \mathcal{E}_z\rangle$ by about 15\% when the duty cycle was increased from 2.5\% to nearly 100\%, i.e. a factor 40 increase in average laser intensity. This indicates that the laser has a measurable effect on the band bending. All measurements reported in the main text were performed with a duty cycle close to 1 in order to maximise the signal-to-noise ratio, and as such they include a small but measurable effect of the laser. For simplicity, photo-ionisation effects are ignored in our modelling (section \ref{sec:BBstudy}), which can be a source of discrepancy when comparing to the experiment.

\begin{figure}[h!]
	\includegraphics{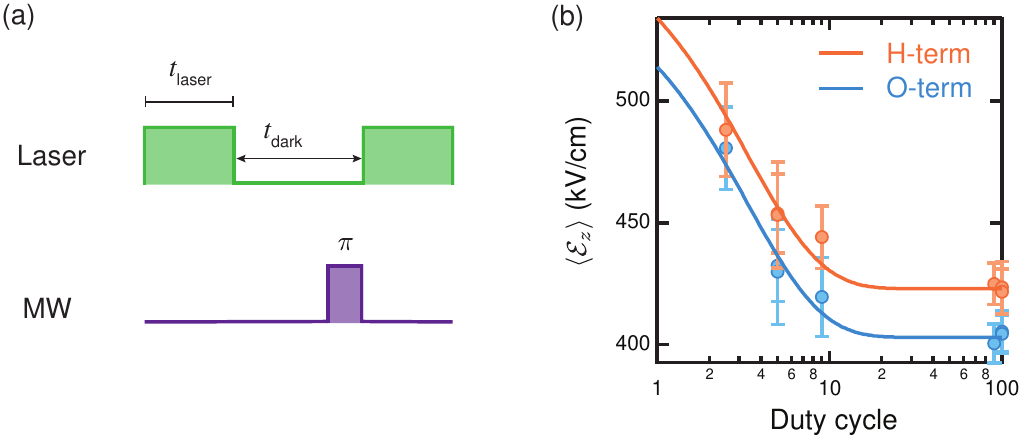}
	\caption{The effect of the laser on the measured electric field. (a) Pulse sequence for ODMR defining the laser pulse time and the dark time. (b) Electric field measurements for oxygen (blue) and hydrogen (orange) terminated diamond for different laser duty cycles, solid line is an exponential fit to the data. }
	\label{Fig: laser}
\end{figure}

\subsection{Electric field vs diamond etching}

Here we investigate the effect of etching the diamond on the electric field. Namely, we used a diamond implanted at $d\approx0-35$~nm (sample \#2) and applied two steps of etching ($\approx15$~nm per step) over partly overlapping patches.

The first etching step was done by removing the TiC/Pt/Au contacts off sample \#2 via acid etching (15 minutes in a boiling mixture of sulphuric acid and sodium nitrate to remove the Pt/Au layers, followed by $5$ minutes ultrasonication in $1$:$1$:$5$ NH$_4$OH\,:\,H$_2$O$_2$\,:\,H$_2$O to remove the TiC), as illustrated in Fig.~\ref{Fig: Etching}a. Following the etching, the diamond was optically clear indicating that there was no metal left, and a recess of $\approx15$~nm was measured by AFM in the regions formerly occupied by the TiC/Pt/Au stack (Figs.\,\ref{Fig: Etching}b,c).

\begin{figure}[tb!]
	\includegraphics[width=0.75\textwidth]{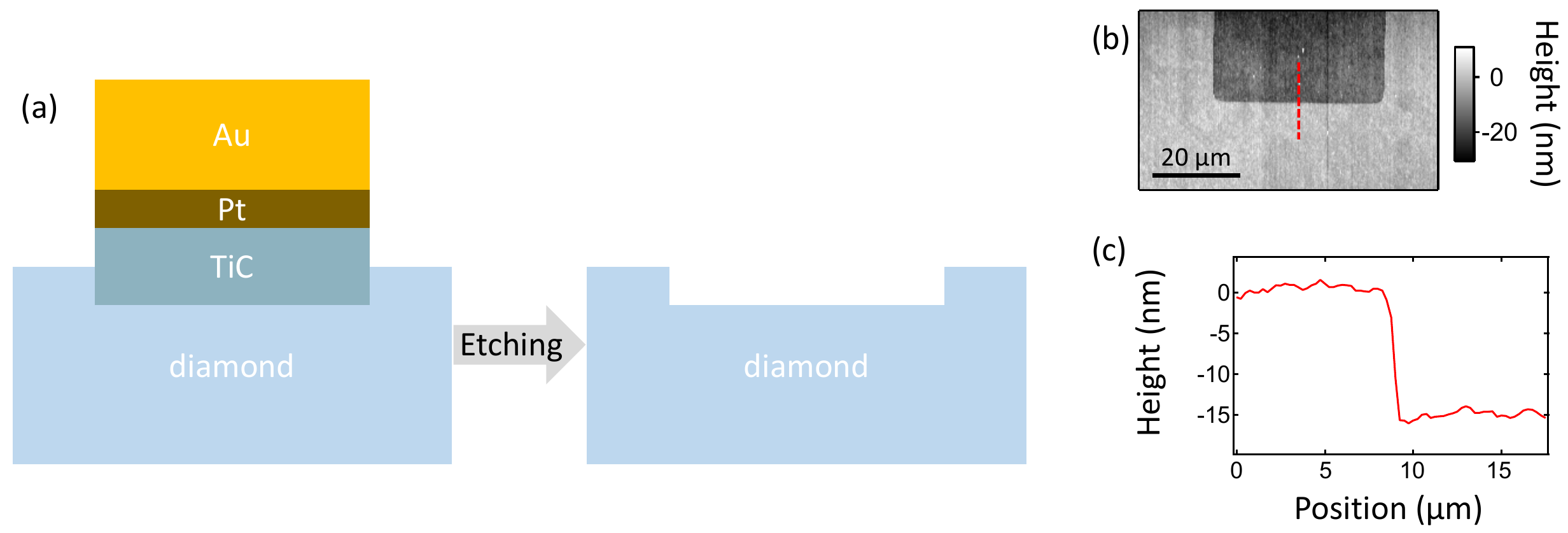}	
	\caption{(a) Illustration of the etching process: the TiC layer formed by annealing is etched away, leaving a recess. (b) Atomic force microscopy (AFM) image of a region thus etched. (c) Line cut extracted from (b) revealing a $\approx15$~nm etching step.}
	\label{Fig: Etching}
\end{figure}

The second etching step was done by reactive ion etching (RIE) through a Au mask, which was subsequently removed via acid etching (15 minutes in a boiling mixture of sulphuric acid and sodium nitrate). An etching step of $\approx15$~nm was measured by AFM and optical profilometry. The etching mask was chosen to partly overlap with the regions etched during the first etching step, so that the diamond eventually had regions that were etched only once ($\approx15$~nm, due to either step 1 or step 2) or etched twice ($\approx30$~nm, due to both steps) as well as non-etched regions (Fig. \ref{Fig: Efield etching}a). Because an acid cleaning was applied as the final step of the process, all the surfaces should be oxygen terminated, yet with possibly different density of surface defects. 

\begin{figure}[b!]
	\includegraphics[width=0.5\textwidth]{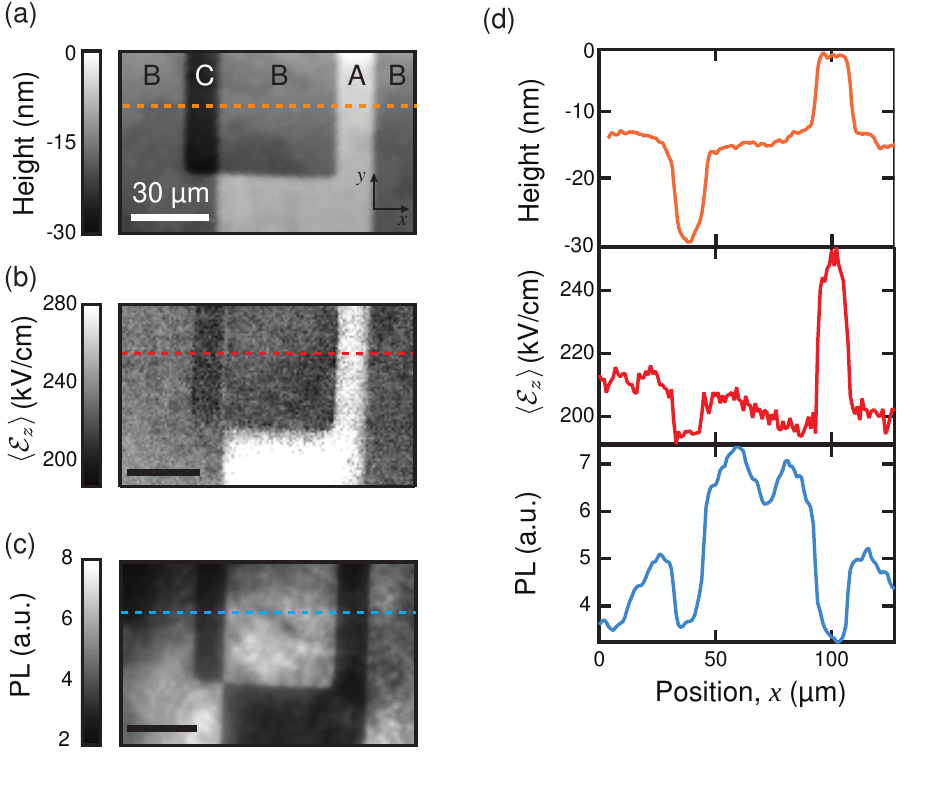}
	\caption{(a) Topography image of the diamond surface obtained by optical profilometry, highlighting (A) a non-etched region, (B) regions etched once and (C) a region etched twice. (b) Electric field map of the same area as in (a).  The plotted range is capped to 280 kV/cm to highlight spatial variations, with actual values reaching up to 320 kV/cm. (c) Corresponding PL image. (d) Line cuts extracted from (a-c), taken along the dashed lines shown in the images.}
	\label{Fig: Efield etching}
\end{figure} 

The topography, electric field, and PL images of a selected area are shown in Figs. \ref{Fig: Efield etching}a-c, respectively, with line cuts shown in Fig. \ref{Fig: Efield etching}d. The electric field is found to be correlated with the surface height, namely the more the diamond was etched, the smaller the electric field became. On the other hand, the PL shows a less trivial correlation, where it is larger in the regions etched once ($\approx15$~nm etching) than in the regions etched twice (30 nm) or not etched at all. 
To interpret these results, let us first consider the effect of the first etching step, which decreased the electric field by $\approx25\%$ and increased the PL by a factor $\approx2$, despite etching away 15 nm off the estimated depth range of the NV centres, $d\approx0-35$~nm. This implies a dramatic reduction in band bending, such that for instance the depletion region for NV$^-$ decreased from 27 nm (i.e. NV$^-$ only at depths $d=27-35$~nm, an 8-nm-thick layer) to just 5 nm (NV$^-$ at $d=20-35$~nm, where $d=0-15$~nm has been etched). The measured $\langle \mathcal{E}_z\rangle$ is averaged over the NV$^-$ depth distribution, which is deeper before the etch hence probes a flatter part of the band bending, leading to only a small decrease in $\langle \mathcal{E}_z\rangle$ (by $\approx25\%$) upon etching even though $\mathcal{E}_z$ is expected to be significantly smaller (by a factor of $~5$ with our example numbers). A possible explanation for such a dramatic reduction in band bending invokes the presence of sub-surface defects present in the non-etched region and induced by the hydrogen plasma initially applied to the sample, which may have resulted in diffusion of hydrogen atoms into the diamond~\cite{Stacey2012}. The etching would then remove these defects to leave only the surface-termination-induced defects as the source of band bending. 

Assuming the second etch results in the same surface as the first etch, with the same density of surface defects, we would expect the PL to decrease relative to the first etch, which it does by a factor $\approx2$. The assumptions made above (5 nm depletion layer and NV depth range $d=0-35$~nm before etching) are clearly over simplistic as they predict no PL at all after a 30 nm etch. Most likely, the implantation profile (at energy 10 keV for this sample) produced a tail that extends deeper than 35 nm. The further decrease in $\langle \mathcal{E}_z\rangle$ upon the second etch can be explained by an increase in the band bending depth (i.e., the length scale of the bending) while the total bending (i.e. the energy offset at the surface) remains unchanged. This is expected as at this stage a significant number of nitrogen defects (donors) has been etched away, leaving a diamond with a lower nitrogen concentration hence a more dilute space charge density. While further work is needed to fully understand the multiple competing effects, these results illustrate the value of in-situ electric field measurements, which provide new insight into band bending. 

\section{Modelling of band bending at the diamond surface} \label{sec:BBstudy}

In this section, we first present our model of the oxygen- and hydrogen-terminated diamond surfaces, then describe how band bending and electric field are calculated within this model, and finally discuss some results.

\subsection{Models of the diamond surface}
\newcommand{\dmax}{d_{\textrm{max}}}
\newcommand{\DSS}{D_{sd}}
\newcommand{\QSA}{Q_{sa}}
\begin{figure}[b!]
	\includegraphics[width=\textwidth]{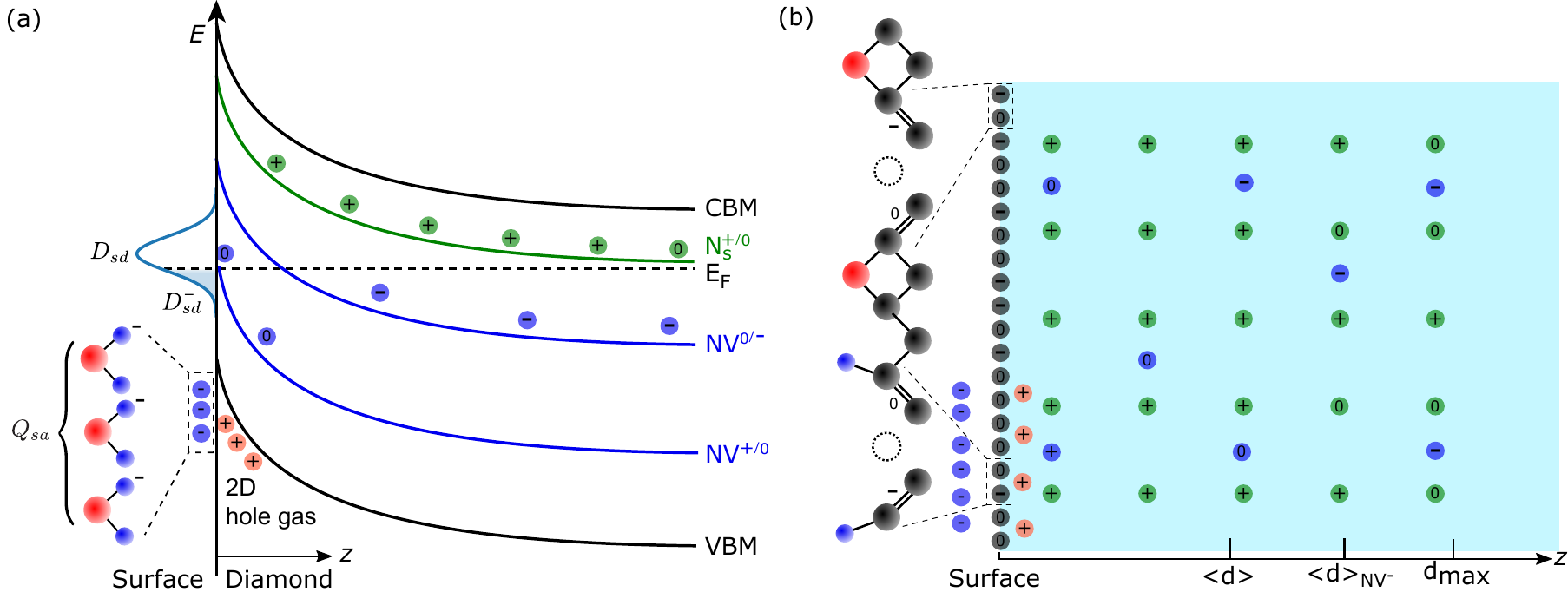}
	\caption{(a) Band diagram of a nitrogen implanted diamond near the surface, showing the conduction band minimum (CBM), valence band maximum (VBM), the N$_{\rm S}^{+/0}$ charging level, the NV$^{0/-}$ charging level, the NV$^{+/0}$ charging level, and the constant Fermi level $E_F$. To the left of the vertical axis are: a depiction of the density of states associated with the \emph{sp$^2$} surface defects, which have a total areal density $\DSS$, and an ionised density $\DSS^-$; a depiction of the adsorbed acceptor molecules present close to the VBM for hydrogen-terminated diamond, with an ionised areal density ($\QSA$), and inducing a 2D hole gas in the diamond. (b) A cartoon depicting the surface and nitrogen induced defects considered in the band modelling (see text).}
	\label{Fig: Diamond Bands}
\end{figure} 

\subsubsection{Oxygen-terminated diamond}

The surface of materials often have electronic energy levels that are distinctly different from the bulk material bands.  In semiconductors, these surface states can exist within the band gap of the material, as depicted in Fig.~\ref{Fig: Diamond Bands}. Any unoccupied (occupied) surface states (with areal density $\DSS$) below (above) the bulk Fermi level will ionise.  However, to maintain charge neutrality within the crystal the total charge of the ionised surface states needs to be compensated by an opposite charge from the bulk. In order for charge to build up in the bulk the bands bend towards the surface, which simultaneously reduces the density of the ionised surface states, until charge neutrality is reached.  Typically the Fermi level at the surface is pinned close to the surface state energy unless a density of donors that fully compensate the density of surface states can be introduced into the material.  In the case of oxygen-terminated diamond we observed a strong electric field which we ascribe to an upward band bending indicative of a high density of surface states.  This is commensurate with published electrical measurements of Fermi level pinning at diamond interfaces~\cite{Zheng2001, Garino2009, Mori1991}.

Information about the density and energy of any surface states present on oxygen terminated surfaces will then be required to model the expected electric field.  The cross section of a (001) diamond surface consists of carbon bonded in a zig-zag pattern between the first and second atomic layer.  Each surface carbon atom has two dangling bonds that are satisfied by bonding to an oxygen termination layer. Oxygen can terminate the diamond in a number of ways: an oxygen atom can bond to two surface carbon atoms in an ether like arrangement (depicted in Fig.~\ref{Fig: Diamond Bands}b), it can double bond to a single surface carbon atom in a ketone like arrangement or a hydroxyl group can terminate a single surface carbon atom. Each of these different arrangements introduce different surface states into the diamond band gap, yet only ether-like terminations are expected to result in upward band bending in nitrogen doped diamond (where the Fermi level is close to the nitrogen donor level at VBM + 3.75 eV) with an unoccupied state at VBM + 3.4 eV~\cite{Sque2006}. Such a high unoccupied energy level is unable to explain the large band bending that was measured on all oxygen terminated samples and thus we don't consider the effect of any oxygen induced surface states on the diamond electronic structure.  Recently a primal defect, the \emph{sp$^2$} defect, of the diamond surface has been identified to have a much lower acceptor energy around VBM + 1.5~eV~\cite{Stacey2018}. The fundamental structure of this defect is shown in Fig.~\ref{Fig: Diamond Bands}b, and occurs when a surface carbon atom is removed from the lattice. The dangling bonds that remain then form double bonds with the exposed sub-surface carbon. As the energy of this defect state is well below the nitrogen donor level, and the density is expected to be large, $\DSS > 10^{13}$~cm$^{-2}$, this defect makes a good candidate to explain the measured electric field under oxygen terminated surfaces. The actual energy of the \emph{sp$^2$} defect surface state is dependent on its local environment, and DFT modelling predict energies within a range from $0.9$~eV to $2.2$~eV.  To capture this range we consider the defect density of states (DOS) to be Gaussian around $1.5$~eV with a full width at half maximum of $0.6$~eV, in agreement with the shape of the resonance peak seen in X-ray absorption experiments probing this defect~\cite{Stacey2018}.

\subsubsection{Bulk defects}

The surface charge is compensated by the dielectric response of the diamond, $\epsilon = 5.8$, the ionisation of any defects within the band gap, predominantly nitrogen donors, and any charge carriers (which is practically zero until the Fermi level is within 1 eV of a diamond band).  The donors within the diamond are formed by the implantation of nitrogen ions, which has a fixed areal density (fluence) of $D_\textrm{areal} = 10^{13}$ nitrogen cm$^{-2}$ in these experiments. For shallow implants with an implant energy below $30$~keV, the depth dependent nitrogen density profile is in general non trivial~\cite{Lehtinen2016} and is particularly sensitive to the angle of incidence, which here is nominally $7^\circ$ from the surface normal but is partly randomised by the roughness of the diamond surface. We chose to approximate the nitrogen density profile by a rectangular function bounded by the surface on one end ($d=0$) and a maximum implant depth parameter, $\dmax$, on the other end. Such a simple function is a reasonably good approximation to both theoretically and experimentally determined profiles in our situation~\cite{Toyli2010, Lehtinen2016,FavaroDeOliveira2015}, and avoids the introduction of multiple parameters. The only parameter is $\dmax$, which can be related to the implant energy via $\dmax = \eta \, E_{\textrm{imp}}$. To maximise the match between the data presented in the main text and theory, a value of $\eta = 3.5$ was chosen, hence a mean depth $\langle d\rangle=\frac{\dmax}{2}=1.75E_{\textrm{imp}}$, also providing a reasonable agreement with previous studies~\cite{Tetienne2018,Lehtinen2016,Toyli2010}. The volumetric density of nitrogen in the diamond is then calculated as $\frac{D_\textrm{areal}}{\dmax}$. The nitrogen implant forms a range of different defects within the diamond: substitutional nitrogen (N$_\textrm{S}$), nitrogen-vacancy centres (NV) and vacancy defects (which can be single, double or multi-vacancy chains). The majority of the implanted nitrogen ends up as N$_\textrm{S}$ with only $\chi \approx 1$\% turning into NV for high nitrogen fluences~\cite{Pezzagna2010}, resulting in densities $D_{\textrm{N}}$ = $(1-\chi)\frac{D_\textrm{areal}}{\dmax}$ and $D_{\textrm{NV}}$ = $\chi\frac{D_\textrm{areal}}{\dmax}$ for the N$_\textrm{S}$ and NV defects, respectively. The N$_\textrm{S}$ acts as a deep acceptor state, $1.7$~eV ($E_{\textrm{N}_{\rm S}} = 3.75$~eV) away from the conduction band.  Whilst the NV defect can become both positively and negatively charged, allowing to act as both a super deep donor ($E_{\textrm{NV}^+} = 0.75-1.05$~eV) and mid gap acceptor ($E_{\textrm{NV}^-} = 2.7-2.8$~eV)~\cite{Deak2014}. For simplicity we do not consider other vacancy related defects. Although the expected density of vacancy defects is larger than the density of nitrogen related defects after implantation, multiple annealing steps at $950^\circ$C and $1200^\circ$C were undertaken to minimise the quantity of vacancy defects and the final density is expected to be low~\cite{Goss2005, Tetienne2018,DeOliveira2017}.

\subsubsection{Hydrogen-terminated diamond}

Having presented our model of the oxygen-terminated surface, we now move on to describe the hydrogen-terminated surface. The clean and ordered hydrogen-terminated diamond surface forms a strong dipole field that reduces the diamond's work function, moving the vacuum energy level below the diamond conduction band (observed as a negative electron affinity~\cite{Takeuchi2005}).  An observed result of this low work function is that it is energetically favourable for electrons in the diamond valence band to be excited to adsorbed acceptor molecules in atmospheric conditions. The total negative charge of ionised adsorbed acceptors ($\QSA$), depicted in Fig.~\ref{Fig: Diamond Bands}a, needs to be compensated by a positive charge within the diamond, and much like under an oxygen-terminated surface the diamond bands bend upwards towards the surface. As this bending is not pinned to an energy level in the diamond band gap, it often extends all the way into the diamond valence band, forming a two-dimensional hole gas confined within $5$~nm of the surface~\cite{RgenRistein2006, Hauf2014} and resulting in the observed surface conductivity of hydrogen-terminated diamonds~\cite{Koslowski2001}. In intrinsic diamond, holes in the valence band are the only positive bulk charge that provides a significant contribution to compensation of the surface charge. In nitrogen doped diamond, however, the ionised nitrogen defects will also partially compensate any surface charge. With a high enough density of nitrogen, the Fermi level at the surface will move far enough from the valence band that no hole layer is formed and the diamond surface ceases to be conductive~\cite{Grotz2012a}. In our samples, however, the primal \emph{sp$^2$} defects should still be present because the hydrogenation is unlikely to remove them~\cite{Stacey2018}. As a consequence, the \emph{sp$^2$} defects will once again pin the surface Fermi level position. Like with oxygen termination, the total charge in the acceptor layer is a function of the Fermi level at the diamond surface~\cite{Newell2016}, however to simplify dealing with the effect of both surface acceptors and ionised \emph{sp$^2$} defects, we consider $\QSA$ to be constant.  We can relate the measured hydrogen ribbon resistivity ($R$) to the density of holes at the diamond surface, $n_h$, using $R = \frac{L}{W\sigma}$, where $L = 100$~$\mu$m and $W = 18$~$\mu$m are the hydrogen ribbon length and width for our devices. The conductivity of the device is given by $\sigma = en_h\mu_h$ where $\mu_h$ is the hole mobility in the ribbon.  Hole mobilites between $\mu_h = 10 - 100$~cm$^{2}$V$^{-1}$s$^{-1}$ are commonly reported~\cite{Sato2013c, Pakes2014}.  Device resistances of our ribbon ranged from initial values of $R \approx 100$~k$\Omega$  to $R > 10$~M$\Omega$ after extensive measurement, not taking into account any contact resistance.  This provides an estimate for the 2D hole density in the range of $n_h = 3.5\times10^{10} - 3.5\times 10^{12}$~cm$^{-2}$. Accounting for the $10$~keV implant in sample \#2 ($\dmax=35$~nm), numerical integration returns that $\QSA$ is within the range $1.08$ to $0.98 \times 10^{13}$~cm$^{-2}$. 

In all cases the upward bending bands will affect the response of the implanted NV defects, de-ionizing the near surface NV where the band bending is strong~\cite{Schreyvogel2016}, and creating a space charge layer of ionised N$_\textrm{S}$ (the green circles in Fig.~\ref{Fig: Diamond Bands}).  The space charge layer creates a significant electric field around the NV$^-$ defects, it is this field that can be observed as ODMR shifts.  To estimate the amount of NV$^-$ and the size of the electric field that the NV$^-$ experience, the shape of this bending needs to be calculated. 

\subsection{Calculating band bending}
\newcommand{\EF}{E_F}
\newcommand{\EV}{E_V}
\newcommand{\EC}{E_C}
\newcommand{\NV}{N_V}
\newcommand{\NC}{N_C}
\newcommand{\mV}{m^*_V}
\newcommand{\mC}{m^*_C}
\newcommand{\ED}{E_D}
\newcommand{\EA}{E_A}
\newcommand{\ND}{N_D}
\newcommand{\NA}{N_A}
\newcommand{\ENS}{E_{\textrm{N}_\textrm{S}}}
\newcommand{\ENVm}{E_{\textrm{NV}^-}}
\newcommand{\ENVp}{E_{\textrm{NV}^+}}
\newcommand{\NNS}{D_{\textrm{N}}}
\newcommand{\NNV}{D_\textrm{NV}}
\newcommand{\QSS}{Q_{sd}}
\newcommand{\QBC}{Q_{BC}}
Poisson's equation provides a classical definition of how a charge distribution $\rho(\vec{r})$ changes the electric potential $V(\vec{r})$ within a dielectric material:
\begin{gather}
\nabla^2 V(\vec{r}) = \frac{-\rho(\vec{r})}{\epsilon\epsilon_0}
\end{gather}
where $\epsilon_0$ is the vacuum permittivity and $\epsilon$ is the relative permittivity of the material. Solving this equation determines how the bands bend given a certain surface potential $V_s$ (or equivalently, a normalised potential $\nu_s$, related to $V_s$ via Eq.~\ref{eq:normalisedpotential}), i.e. the boundary condition at the diamond surface. The shape of the bending is only determined by the charges within the diamond, i.e. if one solves for $\nu_s = 5$ and redefine the surface at $\nu = 3$ discarding the higher potentials, the solution is the same as solving for $\nu_s = 3$~\cite{lüth2014solid}. We take advantage of this and over-solve all our bands with a large $\nu_s$ then redefine the surface at the appropriate potential to compensate a given areal density of charge at the diamond surface.

To determine the charge distribution in the diamond we need to consider the densities of electrons, holes and any ionised defect states (negatively charged acceptors and positively charged donors). The density of electrons, holes and ionised defects is determined by the filling of electronic states in the diamond valence and conduction bands and the defect states, respectively. The filling of electronic states is calculated by convolving the energy dependent DOS with the probability of a state being occupied by an electron at a given energy.  The probability of a state being occupied (unoccupied) is given by the Fermi-Dirac distribution function, 
\begin{equation}
f_D(E) = \frac{1}{1+ \expp{\frac{E-\EF}{kT} }} 
\end{equation}
and its complement, 
\begin{equation}
1-f_D(E) = \frac{1}{1+ \expp{ \frac{\EF - E}{kT} }} 
\end{equation}
where $k$ is the Boltzmann constant, $T$ is the temperature, and the functions are centred around the Fermi level ($\EF$) of a material. The Fermi level can be determined by requiring that the material be charge neutral at the Fermi level, giving
\begin{gather}\label{FERMI_LEVEL}
p + \sum_{\textrm{Donors}}\ND^+ = n +\sum_{\textrm{Acceptors}}\NA^-
\end{gather}
where $p$ ($n$) is the density of holes (electrons), $\ND^+$ is the density of ionised donors, and $\NA^-$ is the density of ionised acceptors in the bulk of diamond in thermal equilibrium. We can calculate the electron and hole densities using the effective mass ($m^*$) approximation for the valence and conduction bands DOS~\cite{StreetmanBanerjee}:
\begin{equation}
\begin{aligned}
\NV(E) &= \frac{(2\mV)}{2\pi^2 h^3}^{3/2}\sqrt{\EV-E}, \\
\NC(E) &= \frac{(2\mC)}{2\pi^2 h^3}^{3/2}\sqrt{E - \EC},
\end{aligned}
\end{equation}
where $h$ is Plank's constant and $\mV$ ($\mC$) is the effective mass of valance (conduction) band. Typically, the Boltzmann approximation of $f_D(E) \approx e^{(\EF-E)/kT}$ for $\EF-E \gg kT$, is used to estimate the electron and hole densities:
\begin{equation}
\begin{aligned}
p(z) &= 2\left(\frac{2\pi \mV kT)}{h^2}\right)^{3/2}e^{\frac{\EV-\EF}{kT}} = \NC e^{\frac{\EV-\EF}{kT}},\\
n(z) &= 2\left(\frac{2\pi \mC kT)}{h^2}\right)^{3/2}e^{\frac{\EF-\EC}{kT}} = \NV e^{\frac{\EF-\EC}{kT}},
\end{aligned}
\end{equation}
where $\NV$ is the effective density of states for holes and $\NC$ for electrons.  In surface transfer doped hydrogen-terminated diamond, the conduction band at the surface crosses through the Fermi level (hence becomes a degenerate semiconductor) with a relatively small carrier density of $2.5\times 10^{12}$~holes cm$^{-2}$, and the Boltzmann approximation is no longer appropriate.  Instead, we have to solve the full convolution:
\begin{equation}
\begin{aligned}
p(z) &= \NV \frac{2}{\sqrt{\pi}}\int_{0}^{\infty}\frac{\sqrt{\epsilon}}{1+e^{\epsilon - \frac{\EV-\EF}{kT}}}\,d\epsilon = \NV \mathcal{F}_{1/2} \left( \frac{\EV-\EF}{kT}\right),\\ 
n(z) &=  \NC\frac{2}{\sqrt{\pi}}\int_{0}^{\infty}\frac{\sqrt{\epsilon}}{1+e^{\epsilon - \frac{\EF - \EC}{kT}}}\,d\epsilon = \NC\mathcal{F}_{1/2}\left(\frac{\EF - \EC}{kT}\right),
\end{aligned}
\end{equation}
where $\mathcal{F}$ is the set of Fermi Dirac integrals, defined as
\begin{equation}
\mathcal{F}_j(\eta) = \int_0^\infty \frac{x^j}{\expp{x-\eta}+ 1}\,dx, \qquad \text{for~} j > -1.
\end{equation}
These integrals can be accurately and rapidly approximated numerically~\cite{Fukushima2015a, Fukushima2015}; in this work the open source python package fdint~$2.02$~\cite{Maddox2017} was used.

To calculate the density of ionised defects, we approximate the defect DOS as a Dirac delta function $\delta(E-E_0)$ around the defect ionisation energy, $E_0=\EA$ for acceptor and $E_0=\ED$ for donor transitions, with defect densities of $\NA$ and $\ND$:
\begin{equation}
\begin{aligned}
\ND^+(z) &= \int_{-\infty}^{\ED(z)}\ND(z)\delta(\ED(z)-E)\,\left[1-f_D(E)\right]\,dE &&= \ND(z)\left[1-f_D(\ED(z))\right], \\
\NA^-(z) &= \int_{\EA(z)}^{\infty}\NA(z)\delta(E-\EA(z))\,f_D(E)\,dE &&= \NA(z)\,f_D(\EA(z)) .
\end{aligned}
\end{equation}
Lateral variations in the surface states and defect densities are ignored and only the charge distribution in $z$ needs to be considered (here $z$ is defined positive going into the diamond). The total charge density within the diamond can then be described by:
\begin{equation}\label{eqn:rhoz}
\rho(\vec{r}) = \rho(z) = ep(z) -en(z) +e\sum_{\textrm{Donors}}\ND^+(z) -e\sum_{\textrm{Acceptors}}\NA^-(z).
\end{equation}

As external potentials affect all the bands equally we re-define the $z$ dependent defect and band energies as a function of a single depth dependent potential. First we define a potential, $\phi(z)$, as the separation between the Fermi level $\EF$ and the intrinsic energy level $E_i$ (i.e. where the Fermi level would be in an intrinsic diamond),  
\begin{equation}\label{Eq: phi potential}
e\phi(z) = \EF - E_i(z)
\end{equation}
where
\begin{equation}
E_i(z) = \frac{1}{2}\EV(z) + \frac{1}{2}\EC(z) -\frac{1}{2}kT\log{\frac{\NC}{\NV}}.
\end{equation}
From Eq.~\ref{Eq: phi potential} the potential $V(z)$ can be defined as how far the bands have shifted from their unbent position,
\begin{gather}
V(z) = \phi(z) - \phi_\textrm{B}
\end{gather}
where
\begin{equation}
\phi_\textrm{B} \equiv \lim_{z\to\infty}\phi(z).
\end{equation}
In order to solve for the band structure it is also useful to define unitless potentials, such that
\begin{equation} \label{eq:normalisedpotential}
\begin{aligned}
u(z) &= \frac{e\phi(z)}{kT} \\
\nu(z) &= \frac{eV(z)}{kT}.
\end{aligned}
\end{equation}
Within the Boltzmann approximation the carrier density can be rewritten in terms of the unitless potential $\nu(z)$:
\begin{equation}
\begin{aligned}
p(z) &\approx \NV \exp\left( \frac{-(\EF-\EV(z))}{kT}\right) \\
&= \NV\exp\left(\frac{-E_i(z) - e\phi(z) + \EV(z)}{kT}\right) \\
&= \NV \exp\left(\frac{-\frac{1}{2}\EV(z) - \frac{1}{2}\EC(z) +\frac{1}{2}kT\log{\frac{\NC}{\NV}} - e\phi(z) + \EV(z)}{kT} \right) \\
&= \NV\sqrt{\frac{\NC}{\NV}} \exp\left( \frac{-e\phi(z)}{kT} \right) \exp\left( \frac{\EV-\EC}{2kT}\right) \\
&= \sqrt{(\NC\NV)}\exp\left(\frac{\EV-\EC}{2kT} \right) \exp\left(-u(z)\right)\\
&= n_i \exp\left(-u(z) \right)  \\
&= n_i \exp\left( -u(\infty) \right) \exp\left(-\nu(z)\right) 
\end{aligned}
\end{equation}
which results in
\begin{equation}
\begin{aligned}
p(z) &= p_\textrm{B}e^{-\nu(z)} \\
n(z) &= n_\textrm{B}e^{\nu(z)}
\end{aligned}
\end{equation}
where $p_\textrm{B}$ and $n_\textrm{B}$ are the hole and electron densities at the Fermi level. These carrier densities can also be express in terms of Fermi-Dirac integrals:
\begin{equation}
\begin{aligned}
p(z) &= \NV\mathcal{F}_{1/2}\left(\frac{\EV-\EF}{kT}-\nu(z)\right), \\
n(z) &= \NC\mathcal{F}_{1/2}\left(\nu(z) + \frac{\EF - \EC}{kT}\right),
\end{aligned}
\end{equation}
which facilitates the use of the aforementioned numerical approximations. 

A donor energy level ($\ED$) represents the energy required to remove an electron from a crystal defect. The following treatment for including bulk defects is general for all donors. The ionised donor density can be defined in terms of the donor energy, such that
\begin{gather}
\ND^+(z) = \ND(1-f_D(\ED(z))) = \ND \frac{1}{1+\expp{\frac{\EF - \ED(z)}{kT}}}.
\end{gather}
Using the unitless potential, this can be rewritten as
\begin{equation}
\ND^+(z) = \ND \frac{1}{1+\expp{\nu(z)}\expp{\frac{\EF-\ED^\textrm{B}}{kT}}}
\end{equation}
where $\ED^\textrm{B}$ is the energy of the donor in the limit  $z\rightarrow \infty$. This is similar for an acceptor level,
\begin{gather}
\NA^-(z) = \NA \frac{1}{1+\expp{-\nu(z)}\expp{\frac{\EA^\textrm{B}-\EF}{kT}}}.
\end{gather}

Using the chain rule, we can analytically reduce the second order Poisson's equation to first order by integration of the charge density with respect to $\nu$:
\begin{align}
\frac{d^2V(z)}{dz^2} &= \frac{kT}{e}\frac{d^2\nu(z)}{dz^2} = \frac{-\rho(z)}{\epsilon\epsilon_0}  \\
\Rightarrow \left(\frac{d\nu}{dz}\right)^2 &= \frac{e}{\epsilon\epsilon_0kT}\int-\rho(z)\,d\nu.
\end{align}
As $\nu(z)$ is unknown, this integral is only solvable if $\rho(z)$ is only a function of $\nu$, and not explicitly a function of $z$. In that case we get:
\begin{gather}\label{}
\int-\rho(z)\,d\nu = \int_0^{\nu(z)}-ep(z) +en(z) -e\sum_{\textrm{Donors}}\ND^+(z) +e\sum_{\textrm{Acceptors}}\NA^-(z)\,d\nu
\end{gather}
By integrating from 0 we ensure that $\frac{d\nu}{dz} = 0$ when $\nu(z) = 0$, i.e. the bands bend until they reach the Fermi level, which occurs when the crystal is charge neutral.

To calculate the Fermi level in the implanted layer, we use the Boltzmann approximation for electron and hole densities and consider the N$_\textrm{S}$ and NV defects. Eq.~\ref{FERMI_LEVEL} becomes:
\begin{equation}\label{FERMI_LEVEL_SPECIFIC}
\begin{aligned}
p + \sum_{\textrm{Donors}}\ND^+ &=
\NC\expp{\frac{\EV-\EF}{kT}} + \frac{\NNS}{1+\expp{\frac{\EF-\ENS}{kT}}} + \frac{\NNV}{1+\expp{\frac{\EF-\ENVp}{kT}}}\\
= n +\sum_{\textrm{Acceptors}}\NA^- &= \NV\expp{\frac{\EF-\EC}{kT}} + \frac{ \NNV}{1+\expp{\frac{\ENVm-\EF}{kT}}}.
\end{aligned}
\end{equation}
By treating $\EF$ as a variable, the roots of Eq.~\ref{FERMI_LEVEL_SPECIFIC} can be determined numerically.

We then solve for $\nu(z)$ by numerical inversion and integration of $\frac{d\nu}{dz}$, giving
\begin{gather}
z(v) = z_0 + \int_{\nu(z_0)}^{\nu}\frac{dz}{d\nu}d\nu
\end{gather}
where $z_0$ is the crystal surface and $\nu(z_0) \equiv \nu_s$ is the potential at the surface, which needs to be determined for the different diamond surfaces.

The surface potential is determined by the requirement for total charge at the surface ($\QSS$) to be equal to the total charge in the bulk $\QBC$, that is
\begin{gather}
\QSS = -\QBC,
\end{gather}
where $\QSS$ can be determined by numerical inversion of $z(\nu)$ and insertion into Eq.~\ref{eqn:rhoz}.  The total surface charge $\QSS$ is determined by the nature of the surface charge.  When we consider just the effects of the \emph{sp$^2$} defect (oxygen-terminated diamond), we calculate the density of ionised surface states ($\DSS^-$) as a function of the potential:

\begin{equation}
\DSS^-(\nu_s) = \int_{-\infty}^{\infty}\frac{\expp{-\left(\dfrac{E-E_s(z)}{2\sigma}\right)^2}}{1+\expp{\dfrac{E-\EF}{kT}}} dE.
\end{equation}
To determine $\QSS$ for an arbitrary $\nu$, this is evaluated such that
\begin{align}
\QSS = \DSS^-(\nu) &= \int_{-\infty}^{\infty}\DSS(E)f_D(E)dE\\ 
&=  \int_{-\infty}^{\infty}\frac{\expp{-\left(\dfrac{\epsilon}{2\sigma}\right)^2}}{1+\expp{-\nu}\expp{\dfrac{\epsilon-\EF+E_s}{kT}}}\,d\epsilon
\end{align}
which requires numerically integration to solve. To determine the bulk charge, we integrate our charge density over $z$ to the maximum implantation depth:
\begin{gather}
\QSS = en_h + e\sum_{\textrm{Donors}} \int_{0}^{\dmax} \ND^+(z)\,dz -e\sum_{\textrm{Acceptors}}\int_{0}^{\dmax}N_{\textrm{A}}^-(z)\,dz .
\end{gather}
Integrating to $\dmax$ ensures that any artificial charge density beyond the max implantation depth (a result of assuming a uniform nitrogen density) has no effect on selecting the correct surface termination. The root of $\left[\QSS(\nu) + \QBC(\nu)\right]_{\nu_S}=0$ determines $\nu_s$, i.e. the surface potential which results in the diamond being charge neutral.  The same approach is taken for hydrogen-terminated diamond except that:
\begin{equation}
\QSS = \DSS^-(\nu) + \QSA
\end{equation}
to account for the charge of the ionised adsorbed acceptor layer.  Finally the two-dimensional density of the hole layer under a given hydrogen termination can be calculated with
\begin{gather}\label{Eqn:Qss_Qbc}
n_h = \int_{0}^{d_{max}}p(z) dz.
\end{gather}

\subsection{Results}

The first step to calculating the band bending is to determine the  Fermi level within the diamond using Eq.~\ref{FERMI_LEVEL_SPECIFIC}. As the nitrogen donor and NV acceptor levels are far from the diamond band edges, the electron and hole densities are practically zero.  This means that the Fermi level is only determined by the conversion ratio of N$_\textrm{S}$ to NV, $\chi$.  Fig.~\ref{FERMIPIC}a plots the left (blue) and right (orange) hand side of Eq.~\ref{FERMI_LEVEL_SPECIFIC} as a function of $\EF$ for $\chi = 1$\% and $20$\% with $\NNS = 10^{18}$~cm$^{-3}$ (larger values of $\NNS$ do not change the point of intersection).  The solutions, where the blue and orange intersect, are plotted in Fig.~\ref{FERMIPIC}b showing the expected bulk Fermi level for a range of NV conversion ratios.  For all modelling in this work, one has $\EF \approx 3.85$~eV.
\begin{figure}[bt!]
	\centering
	\includegraphics{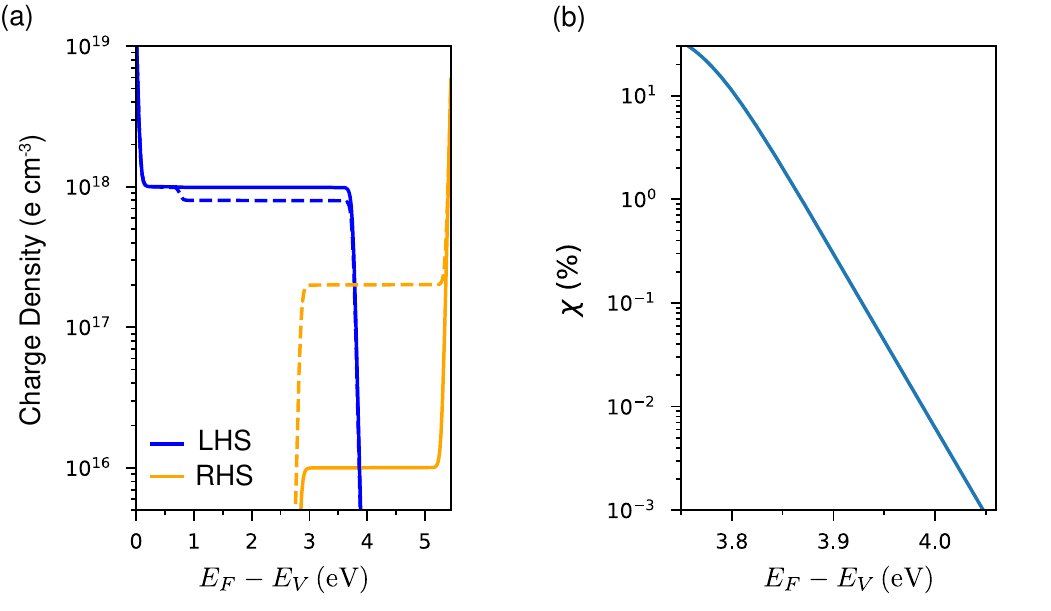}
	\caption{The Fermi level in nitrogen implanted diamond. (a) Visualisation of the neutrality condition (Eq.~\ref{FERMI_LEVEL_SPECIFIC}) showing the left-hand side in blue (LHS, positive charge) and the right-hand side in orange (RHS, negative charge)  as a function of the Fermi level. The N$_\textrm{S}$ to NV conversion efficiency is set to $\chi=1\%$ (solid lines) and $\chi=20\%$ (dashed lines). (b) Calculated Fermi level as a function of $\chi$.}\label{FERMIPIC}
\end{figure}

Fig.~\ref{BAND_BEND_EX}a shows an example of the calculated band bending for two shallow N implants, $D_{\rm areal}=10^{13}$~cm$^{-2}$ with $\dmax=14$~nm ($\sim4$~keV implant) and 70~nm ($\sim20$~keV), at both oxygen and hydrogen surfaces.  For oxygen-terminated surfaces, the integrated DOS, or total number of defect states, is $\DSS=5\times10^{13}$~cm$^{-2}$ or $\approx 2\%$ of the surface~\cite{Stacey2018}.  A total density of ionised surface acceptors $\QSA= 0.98 \times 10^{13}$~cm$^{-2}$ is used to determine the bending under the hydrogen surface.  For each case, a Gaussian DOS is plotted at the surface to depict the surface states, and shading indicates the states that are filled.  In the case of the hydrogen surface with $\dmax=70$~nm, the valence band bends up to the Fermi level generating the conductive 2DHG as expected.  However, for the hydrogen surface with $\dmax=14$~nm where the density of donors is much closer to the surface, the density of surface adsorbed acceptors ($\QSA$) is not large enough to fully compensate the bulk charge and the valence band at the surface no longer bends enough to reach the Fermi level.  The edge of the surface acceptor level then pins the Fermi level, as with oxygen-terminated diamond.  In this case the hydrogen termination is not expected to be conductive, as it is not energetically favourable for a hole layer to form.  It is this transition that is partially responsible for the non-monotonic behaviour of the models displayed in main text Fig. 2a and c, calculated using the best fit value $\QSA = 7\times10^{12}$~cm$^{-2}$. The model then suggests that the sample used in main text Fig. 3 ($\dmax=35$~nm) was close to the conducting/insulating transition. This explains the observed rapid laser induced degradation of the devices, as only small changes to the total acceptor density are required for complete loss of the 2D hole layer near the transition. 

\begin{figure}[b!]
	\centering
	\includegraphics{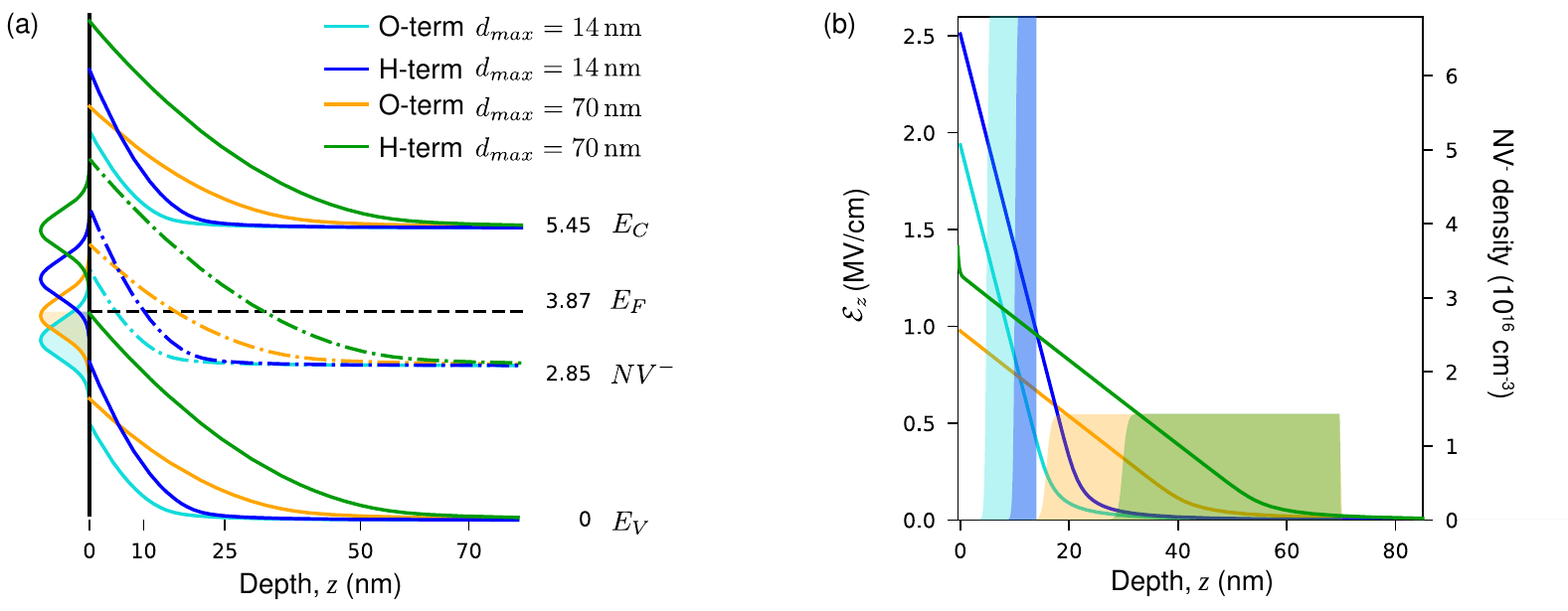}
	\caption{Electronic bands near the surface of a nitrogen implanted diamond. (a) Band position as a function of depth from the surface for both hydrogen and oxygen terminated diamond with different nitrogen implants. (b) Calculated electric field ($\varepsilon_z$) as a function of depth (lines, left axis) and calculated density of NV$^-$ at thermal equilibrium (shaded area, right axis).}\label{BAND_BEND_EX}
\end{figure}

Fig.~\ref{BAND_BEND_EX}b depicts the electric field associated with this bending, and the filled regions show the density of NV$^-$ as determined by $\dmax$ and the ionization state in thermal equilibrium (i.e. where the NV$^-$ band crosses the Fermi level $\EF$ in figure~\ref{BAND_BEND_EX}a).  As the neutral NV$^0$ defect doesn't contribute to the ODMR signal, the measured electric field, denoted as $\langle\varepsilon_z\rangle$, is the average of the field weighted by the density of NV$^-$. In main text Fig. 2a, $\langle\varepsilon_z\rangle$ is plotted as a function of the mean nitrogen implantation depth $\langle d\rangle=\dmax/2$ for a range of surface defect densities $\DSS$, with the inset showing the difference in $\langle\varepsilon_z\rangle$ between H- and O-terminated surface. Main text Fig. 2c was obtained by calculating the relative difference in the total number of NV$^-$ centres at thermal equilibrium between H- and O-terminated surface, as a function of $\langle d\rangle$ and for a range of $\DSS$, assuming the PL is proportional to the number of NV$^-$. Finally, main text Fig. 2d shows $\langle\varepsilon_z\rangle$ versus $\DSS$ for a fixed depth $\dmax=35$~nm. 

We now discuss some of the other predictions enabled by our model. In particular, the de-ionisation of the near surface NV$^-$ due to band bending provides an explanation for the difficulties typically encountered in creating NV$^-$ very close to the surface, especially at $d< 5$~nm~\cite{Pham2015}.  Figure~\ref{model_data}a shows the expected average depth of NV$^-$, $\langle d_{\rm{NV}^-}\rangle$ for a $D_\textrm{areal}=10^{13}$~cm$^{-2}$ nitrogen implant compared to the expected average implant depth, $\langle d\rangle$, which is determined by the implant energy via $\langle d\rangle = \eta E_{\rm{imp}}/2$.  When the surface defect density is equal to or larger than the nitrogen implant density, $\DSS \geqslant D_\textrm{areal}$, $\langle d_{\rm{NV}^-}\rangle \approx 5$~nm  even for $E_{\rm{imp}} = 1$~keV implants expected to produced NVs mainly in the top $2$~nm.  One approach to overcome this issue to produce the nearest surface NV$^-$ possible is to increase the implant density.  Fig~\ref{model_data}c shows the total NV ionisation ratio as a function of $D_\textrm{areal}$ with a relatively low surface defect density of $\DSS = 10^{13}$~cm$^{-2}$.  Even for the $E_{\rm{imp}}=3$~keV implant, $D_\textrm{areal} > \DSS$ is required to achieve complete NV$^-$ ionisation and produce near-surface NV$^-$.  This problem is compounded by the expected high density of surface defects~\cite{Stacey2018}, highlighting the importance of controlling the formation of this defect for near-surface sensing.

Figure~\ref{model_data}c shows the relative population of substitutional nitrogen in its neutral state (N$_{\rm{S}}^0$), known as the P1 centre in electron spin resonance measurements, as a function of implantation energy for a  $D_\textrm{areal}$ in a diamond with a surface defect density $\DSS = 10^{13}$~cm$^{-2}$.  The concentration of P1 is a function of both the implant energy and the density of surface states.  No P1 centres are expected to exist in low energy implants, the energy of which is $\DSS$ dependent potentially providing an independent way to verify the density of surface defects in a sample.

\begin{figure}[t!]
    \centering
    \includegraphics{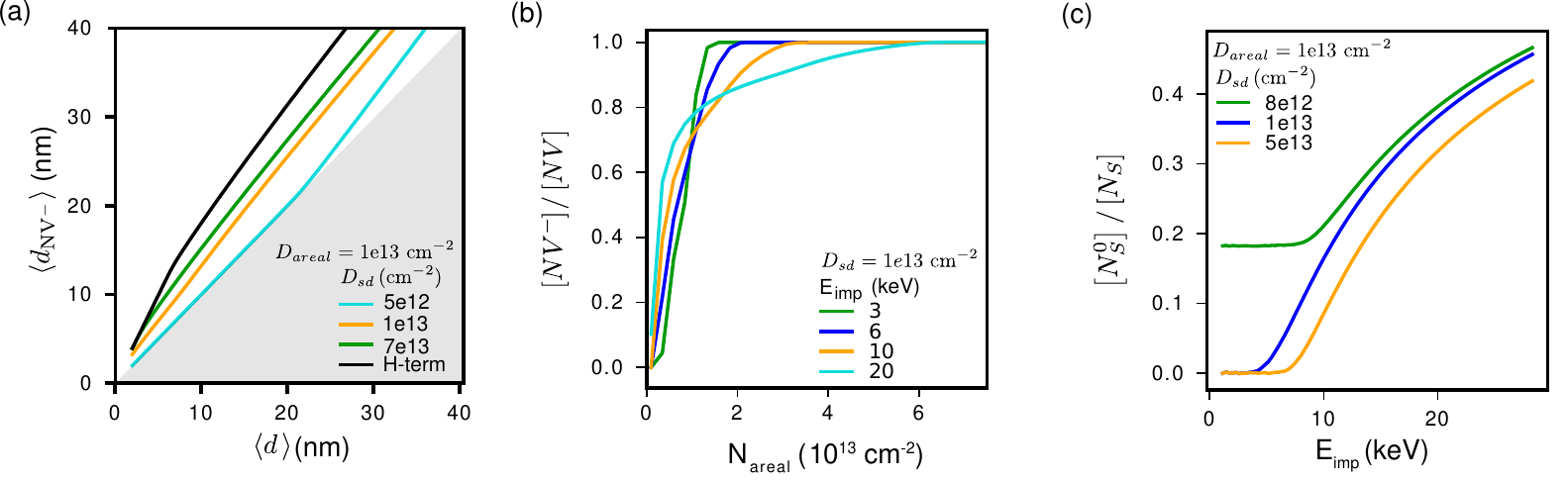}
    \caption{(a) The average NV$^-$ depth versus the average nitrogen depth from implantation for a range of different $\DSS$ densities, and for a hydrogen-terminated diamond (black line), with a fixed implant density of $D_\textrm{areal}=10^{13}$~cm$^{-2}$. (b) The ratio of NV$^-$ to total NV as a function of implantation density for common implantation energies. (c) The ratio of unionised N$_\textrm{S}$, also known as P1 centres, as a function of implantation energy for an implant density of $D_\textrm{areal}=10^{13}$~cm$^{-2}$.}\label{model_data}
\end{figure}

\end{widetext}

\bibliographystyle{naturemag}
\bibliography{bib}

\begin{thebibliography}{10}
\expandafter\ifx\csname url\endcsname\relax
  \def\url#1{\texttt{#1}}\fi
\expandafter\ifx\csname urlprefix\endcsname\relax\def\urlprefix{URL }\fi
\providecommand{\bibinfo}[2]{#2}
\providecommand{\eprint}[2][]{\url{#2}}

\bibitem{Zhang2012}
\bibinfo{author}{Zhang, Z.} \& \bibinfo{author}{Yates, J.~T.}
\newblock \bibinfo{title}{{Band bending in semiconductors: Chemical and
  physical consequences at surfaces and interfaces}}.
\newblock \emph{\bibinfo{journal}{Chem. Rev.}} \textbf{\bibinfo{volume}{112}},
  \bibinfo{pages}{5520--5551} (\bibinfo{year}{2012}).

\bibitem{Kotadiya2018}
\bibinfo{author}{Kotadiya, N.~B.} \emph{et~al.}
\newblock \bibinfo{title}{Universal strategy for ohmic hole injection into
  organic semiconductors with high ionization energies}.
\newblock \emph{\bibinfo{journal}{Nature materials}}
  \textbf{\bibinfo{volume}{17}}, \bibinfo{pages}{329} (\bibinfo{year}{2018}).

\bibitem{Simon2010}
\bibinfo{author}{Simon, J.}, \bibinfo{author}{Protasenko, V.},
  \bibinfo{author}{Lian, C.}, \bibinfo{author}{Xing, H.} \&
  \bibinfo{author}{Jena, D.}
\newblock \bibinfo{title}{Polarization-induced hole doping in
  wide{\textendash}band-gap uniaxial semiconductor heterostructures}.
\newblock \emph{\bibinfo{journal}{Science}} \textbf{\bibinfo{volume}{327}},
  \bibinfo{pages}{60--64} (\bibinfo{year}{2010}).

\bibitem{Stathis2006}
\bibinfo{author}{Stathis, J.} \& \bibinfo{author}{Zafar, S.}
\newblock \bibinfo{title}{The negative bias temperature instability in mos
  devices: A review}.
\newblock \emph{\bibinfo{journal}{Microelectronics Reliability}}
  \textbf{\bibinfo{volume}{46}}, \bibinfo{pages}{270 -- 286}
  (\bibinfo{year}{2006}).

\bibitem{Zhang2006}
\bibinfo{author}{Zhang, P.} \emph{et~al.}
\newblock \bibinfo{title}{Electronic transport in nanometre-scale
  silicon-on-insulator membranes}.
\newblock \emph{\bibinfo{journal}{Nature}} \textbf{\bibinfo{volume}{439}},
  \bibinfo{pages}{703--706} (\bibinfo{year}{2006}).

\bibitem{Kaczr2018}
\bibinfo{author}{Kaczer, B.} \emph{et~al.}
\newblock \bibinfo{title}{A brief overview of gate oxide defect properties and
  their relation to mosfet instabilities and device and circuit time-dependent
  variability}.
\newblock \emph{\bibinfo{journal}{Microelectronics Reliability}}
  \textbf{\bibinfo{volume}{81}}, \bibinfo{pages}{186 -- 194}
  (\bibinfo{year}{2018}).

\bibitem{Weber2010}
\bibinfo{author}{Weber, J.~R.} \emph{et~al.}
\newblock \bibinfo{title}{{Quantum computing with defects.}}
\newblock \emph{\bibinfo{journal}{Proceedings of the National Academy of
  Sciences of the United States of America}} \textbf{\bibinfo{volume}{107}},
  \bibinfo{pages}{8513--8518} (\bibinfo{year}{2010}).

\bibitem{Kaviani2014}
\bibinfo{author}{Kaviani, M.} \emph{et~al.}
\newblock \bibinfo{title}{Proper surface termination for luminescent
  near-surface nv centers in diamond}.
\newblock \emph{\bibinfo{journal}{Nano Letters}} \textbf{\bibinfo{volume}{14}},
  \bibinfo{pages}{4772--4777} (\bibinfo{year}{2014}).

\bibitem{Usman2016}
\bibinfo{author}{Usman, M.} \emph{et~al.}
\newblock \bibinfo{title}{{Spatial metrology of dopants in silicon with exact
  lattice site precision}}.
\newblock \emph{\bibinfo{journal}{Nature Nanotechnology}}
  \textbf{\bibinfo{volume}{11}}, \bibinfo{pages}{763} (\bibinfo{year}{2016}).

\bibitem{Kronik1999}
\bibinfo{author}{Kronik, L.} \& \bibinfo{author}{Shapira, Y.}
\newblock \bibinfo{title}{Surface photovoltage phenomena: theory, experiment,
  and applications}.
\newblock \emph{\bibinfo{journal}{Surface Science Reports}}
  \textbf{\bibinfo{volume}{37}}, \bibinfo{pages}{1 -- 206}
  (\bibinfo{year}{1999}).

\bibitem{Ishii2004}
\bibinfo{author}{Ishii, H.} \emph{et~al.}
\newblock \bibinfo{title}{Kelvin probe study of band bending at organic
  semiconductor/metal interfaces: examination of fermi level alignment}.
\newblock \emph{\bibinfo{journal}{physica status solidi (a)}}
  \textbf{\bibinfo{volume}{201}}, \bibinfo{pages}{1075--1094}
  (\bibinfo{year}{2004}).

\bibitem{Butler2014}
\bibinfo{author}{Butler, C.~J.} \emph{et~al.}
\newblock \bibinfo{title}{Mapping polarization induced surface band bending on
  the rashba semiconductor bitei}.
\newblock \emph{\bibinfo{journal}{Nature Communications}}
  \textbf{\bibinfo{volume}{5}}, \bibinfo{pages}{4066} (\bibinfo{year}{2014}).

\bibitem{Doherty2013}
\bibinfo{author}{Doherty, M.~W.} \emph{et~al.}
\newblock \bibinfo{title}{{The nitrogen-vacancy colour centre in diamond}}.
\newblock \emph{\bibinfo{journal}{Physics Reports}}
  \textbf{\bibinfo{volume}{528}}, \bibinfo{pages}{1--45}
  (\bibinfo{year}{2013}).

\bibitem{Dolde2011}
\bibinfo{author}{Dolde, F.} \emph{et~al.}
\newblock \bibinfo{title}{{Electric-field sensing using single diamond spins}}.
\newblock \emph{\bibinfo{journal}{Nature Physics}}
  \textbf{\bibinfo{volume}{7}}, \bibinfo{pages}{459--463}
  (\bibinfo{year}{2011}).

\bibitem{Strobel2004}
\bibinfo{author}{Strobel, P.}, \bibinfo{author}{Riedel, M.},
  \bibinfo{author}{Ristein, J.} \& \bibinfo{author}{Ley, L.}
\newblock \bibinfo{title}{Surface transfer doping of diamond}.
\newblock \emph{\bibinfo{journal}{Nature}} \textbf{\bibinfo{volume}{430}},
  \bibinfo{pages}{439} (\bibinfo{year}{2004}).

\bibitem{Tetienne2017}
\bibinfo{author}{Tetienne, J.-P.} \emph{et~al.}
\newblock \bibinfo{title}{{Quantum imaging of current flow in graphene}}.
\newblock \emph{\bibinfo{journal}{Science Advances}}
  \textbf{\bibinfo{volume}{3}}, \bibinfo{pages}{e1602429}
  (\bibinfo{year}{2017}).

\bibitem{Degen2017}
\bibinfo{author}{Degen, C.~L.}, \bibinfo{author}{Reinhard, F.} \&
  \bibinfo{author}{Cappellaro, P.}
\newblock \bibinfo{title}{{Quantum sensing}}.
\newblock \emph{\bibinfo{journal}{Reviews of Modern Physics}}
  \textbf{\bibinfo{volume}{89}}, \bibinfo{pages}{035002}
  (\bibinfo{year}{2017}).

\bibitem{Rondin2014}
\bibinfo{author}{Rondin, L.} \emph{et~al.}
\newblock \bibinfo{title}{{Magnetometry with nitrogen-vacancy defects in
  diamond}}.
\newblock \emph{\bibinfo{journal}{Reports on Progress in Physics}}
  \textbf{\bibinfo{volume}{77}}, \bibinfo{pages}{056503}
  (\bibinfo{year}{2014}).

\bibitem{Casola2018}
\bibinfo{author}{Casola, F.}, \bibinfo{author}{{Van Der Sar}, T.} \&
  \bibinfo{author}{Yacoby, A.}
\newblock \bibinfo{title}{{Probing condensed matter physics with magnetometry
  based on nitrogen-vacancy centres in diamond}}.
\newblock \emph{\bibinfo{journal}{Nature Reviews Materials}}
  \textbf{\bibinfo{volume}{3}}, \bibinfo{pages}{17088} (\bibinfo{year}{2018}).

\bibitem{Iwasaki2017}
\bibinfo{author}{Iwasaki, T.} \emph{et~al.}
\newblock \bibinfo{title}{{Direct Nanoscale Sensing of the Internal Electric
  Field in Operating Semiconductor Devices Using Single Electron Spins}}.
\newblock \emph{\bibinfo{journal}{ACS Nano}} \textbf{\bibinfo{volume}{11}},
  \bibinfo{pages}{1238--1245} (\bibinfo{year}{2017}).

\bibitem{Dolde2014}
\bibinfo{author}{Dolde, F.} \emph{et~al.}
\newblock \bibinfo{title}{{Nanoscale detection of a single fundamental charge
  in ambient conditions using the NV - Center in diamond}}.
\newblock \emph{\bibinfo{journal}{Physical Review Letters}}
  \textbf{\bibinfo{volume}{112}}, \bibinfo{pages}{097603}
  (\bibinfo{year}{2014}).

\bibitem{Zhang2018}
\bibinfo{author}{{Zhang}, Q.} \emph{et~al.}
\newblock \bibinfo{title}{{Single rare-earth ions as atomic-scale probes in
  ultra-scaled transistors}}.
\newblock \emph{\bibinfo{journal}{Preprint}} \bibinfo{pages}{arXiv:1803.01573}
  (\bibinfo{year}{2018}).

\bibitem{Falk2014}
\bibinfo{author}{Falk, A.~L.} \emph{et~al.}
\newblock \bibinfo{title}{{Electrically and mechanically tunable electron spins
  in silicon carbide color centers}}.
\newblock \emph{\bibinfo{journal}{Physical Review Letters}}
  \textbf{\bibinfo{volume}{112}}, \bibinfo{pages}{187601}
  (\bibinfo{year}{2014}).

\bibitem{Ohno2012}
\bibinfo{author}{Ohno, K.} \emph{et~al.}
\newblock \bibinfo{title}{Engineering shallow spins in diamond with nitrogen
  delta-doping}.
\newblock \emph{\bibinfo{journal}{Applied Physics Letters}}
  \textbf{\bibinfo{volume}{101}}, \bibinfo{pages}{082413}
  (\bibinfo{year}{2012}).

\bibitem{Lesik2016}
\bibinfo{author}{Lesik, M.} \emph{et~al.}
\newblock \bibinfo{title}{Production of bulk nv centre arrays by shallow
  implantation and diamond cvd overgrowth}.
\newblock \emph{\bibinfo{journal}{physica status solidi (a)}}
  \textbf{\bibinfo{volume}{213}}, \bibinfo{pages}{2594--2600}
  (\bibinfo{year}{2016}).

\bibitem{Stacey2018}
\bibinfo{author}{Stacey, A.} \emph{et~al.}
\newblock \bibinfo{title}{{Evidence for primal $sp^2$ defects at the diamond
  surface: candidates for electron trapping and noise sources}}.
\newblock \emph{\bibinfo{journal}{Preprint}} \bibinfo{pages}{arXiv:1807.02946}
  (\bibinfo{year}{2018}).

\bibitem{Lehtinen2016}
\bibinfo{author}{Lehtinen, O.} \emph{et~al.}
\newblock \bibinfo{title}{{Molecular dynamics simulations of shallow nitrogen
  and silicon implantation into diamond}}.
\newblock \emph{\bibinfo{journal}{Phys. Rev. B}} \textbf{\bibinfo{volume}{93}},
  \bibinfo{pages}{35202} (\bibinfo{year}{2016}).

\bibitem{Doherty2012}
\bibinfo{author}{Doherty, M.~W.} \emph{et~al.}
\newblock \bibinfo{title}{{Theory of the ground-state spin of the NV$^-$ center
  in diamond}}.
\newblock \emph{\bibinfo{journal}{Physical Review B}}
  \textbf{\bibinfo{volume}{85}}, \bibinfo{pages}{205203}
  (\bibinfo{year}{2012}).

\bibitem{Pakes2014}
\bibinfo{author}{Pakes, C.~I.}, \bibinfo{author}{Garrido, J.~a.} \&
  \bibinfo{author}{Kawarada, H.}
\newblock \bibinfo{title}{{Diamond surface conductivity: Properties, devices,
  and sensors}}.
\newblock \emph{\bibinfo{journal}{MRS Bulletin}} \textbf{\bibinfo{volume}{39}},
  \bibinfo{pages}{542--548} (\bibinfo{year}{2014}).

\bibitem{Hauf2011}
\bibinfo{author}{Hauf, M.~V.} \emph{et~al.}
\newblock \bibinfo{title}{{Chemical control of the charge state of
  nitrogen-vacancy centers in diamond}}.
\newblock \emph{\bibinfo{journal}{Physical Review B}}
  \textbf{\bibinfo{volume}{83}}, \bibinfo{pages}{081304}
  (\bibinfo{year}{2011}).

\bibitem{Dhomkar2018}
\bibinfo{author}{Dhomkar, S.}, \bibinfo{author}{Jayakumar, H.},
  \bibinfo{author}{Zangara, P.~R.} \& \bibinfo{author}{Meriles, C.~A.}
\newblock \bibinfo{title}{Charge dynamics in near-surface, variable-density
  ensembles of nitrogen-vacancy centers in diamond}.
\newblock \emph{\bibinfo{journal}{Nano Letters}} \textbf{\bibinfo{volume}{18}},
  \bibinfo{pages}{4046--4052} (\bibinfo{year}{2018}).

\bibitem{DeOliveira2017}
\bibinfo{author}{de~Oliveira, F.~F.} \emph{et~al.}
\newblock \bibinfo{title}{{Tailoring spin defects in diamond by lattice
  charging}}.
\newblock \emph{\bibinfo{journal}{Nat. Commun.}} \textbf{\bibinfo{volume}{8}},
  \bibinfo{pages}{15409} (\bibinfo{year}{2017}).

\bibitem{Kolkowitz2015}
\bibinfo{author}{Kolkowitz, S.} \emph{et~al.}
\newblock \bibinfo{title}{{Probing Johnson noise and ballistic transport in
  normal metals with a single-spin qubit}}.
\newblock \emph{\bibinfo{journal}{Science}} \textbf{\bibinfo{volume}{347}},
  \bibinfo{pages}{1129} (\bibinfo{year}{2015}).

\bibitem{Tetienne2018}
\bibinfo{author}{Tetienne, J.-P.} \emph{et~al.}
\newblock \bibinfo{title}{{Spin properties of dense near-surface ensembles of
  nitrogen-vacancy centers in diamond}}.
\newblock \emph{\bibinfo{journal}{Physical Review B}}
  \textbf{\bibinfo{volume}{97}}, \bibinfo{pages}{085402}
  (\bibinfo{year}{2018}).

\bibitem{Lovchinsky2016}
\bibinfo{author}{Lovchinsky, I.} \emph{et~al.}
\newblock \bibinfo{title}{Nuclear magnetic resonance detection and spectroscopy
  of single proteins using quantum logic}.
\newblock \emph{\bibinfo{journal}{Science}} \textbf{\bibinfo{volume}{351}},
  \bibinfo{pages}{836--841} (\bibinfo{year}{2016}).

\bibitem{Kim2015}
\bibinfo{author}{Kim, C.~K.} \emph{et~al.}
\newblock \bibinfo{title}{{Temperature control for the gate workfunction
  engineering of TiC film by atomic layer deposition}}.
\newblock \emph{\bibinfo{journal}{Solid. State. Electron.}}
  \textbf{\bibinfo{volume}{114}}, \bibinfo{pages}{90--93}
  (\bibinfo{year}{2015}).

\bibitem{Hauf2014}
\bibinfo{author}{Hauf, M.~V.} \emph{et~al.}
\newblock \bibinfo{title}{{Low dimensionality of the surface conductivity of
  diamond}}.
\newblock \emph{\bibinfo{journal}{Physical Review B}}
  \textbf{\bibinfo{volume}{89}}, \bibinfo{pages}{115426}
  (\bibinfo{year}{2014}).

\bibitem{Akhgar2016}
\bibinfo{author}{Akhgar, G.} \emph{et~al.}
\newblock \bibinfo{title}{{Strong and Tunable Spin-Orbit Coupling in a
  Two-Dimensional Hole Gas in Ionic-Liquid Gated Diamond Devices}}.
\newblock \emph{\bibinfo{journal}{Nano Letters}} \textbf{\bibinfo{volume}{16}},
  \bibinfo{pages}{3768--3773} (\bibinfo{year}{2016}).

\bibitem{Simpson2016}
\bibinfo{author}{Simpson, D.~A.} \emph{et~al.}
\newblock \bibinfo{title}{{Magneto-optical imaging of thin magnetic films using
  spins in diamond}}.
\newblock \emph{\bibinfo{journal}{Sci. Rep.}} \textbf{\bibinfo{volume}{6}},
  \bibinfo{pages}{22797} (\bibinfo{year}{2016}).

\bibitem{Doherty2014}
\bibinfo{author}{Doherty, M.~W.} \emph{et~al.}
\newblock \bibinfo{title}{{Measuring the defect structure orientation of a
  single NV- centre in diamond}}.
\newblock \emph{\bibinfo{journal}{New Journal of Physics}}
  \textbf{\bibinfo{volume}{16}}, \bibinfo{pages}{0--20} (\bibinfo{year}{2014}).

\bibitem{Michl2014}
\bibinfo{author}{Michl, J.} \emph{et~al.}
\newblock \bibinfo{title}{{Perfect alignment and preferential orientation of
  nitrogen-vacancy centers during chemical vapor deposition diamond growth on
  (111) surfaces}}.
\newblock \emph{\bibinfo{journal}{Applied Physics Letters}}
  \textbf{\bibinfo{volume}{104}}, \bibinfo{pages}{102407}
  (\bibinfo{year}{2014}).

\bibitem{Felton2009}
\bibinfo{author}{Felton, S.} \emph{et~al.}
\newblock \bibinfo{title}{{Hyperfine interaction in the ground state of the
  negatively charged nitrogen vacancy center in diamond}}.
\newblock \emph{\bibinfo{journal}{Physical Review B}}
  \textbf{\bibinfo{volume}{79}}, \bibinfo{pages}{075203}
  (\bibinfo{year}{2009}).

\bibitem{Lesik2014}
\bibinfo{author}{Lesik, M.} \emph{et~al.}
\newblock \bibinfo{title}{{Perfect preferential orientation of nitrogen-vacancy
  defects in a synthetic diamond sample}}.
\newblock \emph{\bibinfo{journal}{Applied Physics Letters}}
  \textbf{\bibinfo{volume}{104}}, \bibinfo{pages}{113107}
  (\bibinfo{year}{2014}).

\bibitem{Stoneham1968}
\bibinfo{author}{Stoneham, A.~M.}
\newblock \bibinfo{title}{{The shapes of inhomogeneously broadened resonance
  lines II. Second-order effects}}.
\newblock \emph{\bibinfo{journal}{J. Phys. C Solid State Phys.}}
  \textbf{\bibinfo{volume}{1}}, \bibinfo{pages}{302} (\bibinfo{year}{1968}).

\bibitem{Love2011}
\bibinfo{author}{Love, A. E.~H.}
\newblock \emph{\bibinfo{title}{{A treatise on the mathematical theory of
  elasticity}}} (\bibinfo{publisher}{Dover publications}, \bibinfo{address}{New
  york}, \bibinfo{year}{2011}).

\bibitem{Lang1991}
\bibinfo{author}{Lang, A.~R.}, \bibinfo{author}{Moore, M.},
  \bibinfo{author}{Makepeace, A. P.~W.}, \bibinfo{author}{Wierzchowski, W.} \&
  \bibinfo{author}{Welbourn, C.~M.}
\newblock \bibinfo{title}{{On the Dilatation of Synthetic Type Ib Diamond by
  Substitutional Nitrogen Impurity}}.
\newblock \emph{\bibinfo{journal}{Philos. Trans. R. Soc. A Math. Phys. Eng.
  Sci.}} \textbf{\bibinfo{volume}{337}}, \bibinfo{pages}{497--520}
  (\bibinfo{year}{1991}).

\bibitem{Biktagirov2017}
\bibinfo{author}{Biktagirov, T.~B.} \emph{et~al.}
\newblock \bibinfo{title}{{Strain broadening of the 1042-nm zero phonon line of
  the NV- center in diamond: A promising spectroscopic tool for defect
  tomography}}.
\newblock \emph{\bibinfo{journal}{Phys. Rev. B}} \textbf{\bibinfo{volume}{96}},
  \bibinfo{pages}{075205} (\bibinfo{year}{2017}).

\bibitem{Nesladek1998}
\bibinfo{author}{Nesl{\'{a}}dek, M.} \emph{et~al.}
\newblock \bibinfo{title}{{Dominant defect levels in diamond thin films: A
  photocurrent and electron paramagnetic resonance study}}.
\newblock \emph{\bibinfo{journal}{Appl. Phys. Lett.}}
  \textbf{\bibinfo{volume}{72}}, \bibinfo{pages}{3306--3308}
  (\bibinfo{year}{1998}).

\bibitem{Ershov1998}
\bibinfo{author}{Ershov, M.} \emph{et~al.}
\newblock \bibinfo{title}{{Negative capacitance effect in semiconductor
  devices}}.
\newblock \emph{\bibinfo{journal}{IEEE Transactions on Electron Devices}}
  \textbf{\bibinfo{volume}{45}}, \bibinfo{pages}{2196--2206}
  (\bibinfo{year}{1998}).
\newblock \eprint{9806145}.

\bibitem{Jonscher1986}
\bibinfo{author}{Jonscher, A.~K.}
\newblock \bibinfo{title}{{The physical origin of negative capacitance}}.
\newblock \emph{\bibinfo{journal}{J. Chem. Soc. Faraday Trans. 2}}
  \textbf{\bibinfo{volume}{82}}, \bibinfo{pages}{75} (\bibinfo{year}{1986}).

\bibitem{Lemmi1999}
\bibinfo{author}{Lemmi, F.} \& \bibinfo{author}{Johnson, N.~M.}
\newblock \bibinfo{title}{{Negative capacitance in forward biased hydrogenated
  amorphous silicon diodes}}.
\newblock \emph{\bibinfo{journal}{Appl. Phys. Lett. Journal of Applied
  Physics}} \textbf{\bibinfo{volume}{74}}, \bibinfo{pages}{251--2845}
  (\bibinfo{year}{1999}).

\bibitem{Leroy2006}
\bibinfo{author}{Leroy, W.~P.}, \bibinfo{author}{Detavernier, C.},
  \bibinfo{author}{{Van Meirhaeghe}, R.~L.}, \bibinfo{author}{Kellock, A.~J.}
  \& \bibinfo{author}{Lavoie, C.}
\newblock \bibinfo{title}{{Solid-state formation of titanium carbide and
  molybdenum carbide as contacts for carbon-containing semiconductors}}.
\newblock \emph{\bibinfo{journal}{J. Appl. Phys.}}
  \textbf{\bibinfo{volume}{99}}, \bibinfo{pages}{063704}
  (\bibinfo{year}{2006}).

\bibitem{Stacey2012}
\bibinfo{author}{Stacey, A.} \emph{et~al.}
\newblock \bibinfo{title}{{Depletion of nitrogen-vacancy color centers in
  diamond via hydrogen passivation}}.
\newblock \emph{\bibinfo{journal}{Applied Physics Letters}}
  \textbf{\bibinfo{volume}{100}}, \bibinfo{pages}{6--10}
  (\bibinfo{year}{2012}).

\bibitem{Zheng2001}
\bibinfo{author}{Zheng, J.~C.}, \bibinfo{author}{Xie, X.~N.},
  \bibinfo{author}{Wee, A.~T.} \& \bibinfo{author}{Loh, K.~P.}
\newblock \bibinfo{title}{{Oxygen-induced surface state on diamond (100)}}.
\newblock \emph{\bibinfo{journal}{Diam. Relat. Mater.}}
  \textbf{\bibinfo{volume}{10}}, \bibinfo{pages}{500--505}
  (\bibinfo{year}{2001}).

\bibitem{Garino2009}
\bibinfo{author}{Garino, Y.}, \bibinfo{author}{Teraji, T.},
  \bibinfo{author}{Koizumi, S.}, \bibinfo{author}{Koide, Y.} \&
  \bibinfo{author}{Ito, T.}
\newblock \bibinfo{title}{{P-type diamond Schottky diodes fabricated by vacuum
  ultraviolet light/ozone surface oxidation: Comparison with diodes based on
  wet-chemical oxidation}}.
\newblock \emph{\bibinfo{journal}{Phys. Status Solidi Appl. Mater. Sci.}}
  \textbf{\bibinfo{volume}{206}}, \bibinfo{pages}{2082--2085}
  (\bibinfo{year}{2009}).

\bibitem{Mori1991}
\bibinfo{author}{Mori, Y.}, \bibinfo{author}{Kawarada, H.} \&
  \bibinfo{author}{Hiraki, A.}
\newblock \bibinfo{title}{{Properties of metal/diamond interfaces and effects
  of oxygen adsorbed onto diamond surface}}.
\newblock \emph{\bibinfo{journal}{Appl. Phys. Lett.}}
  \textbf{\bibinfo{volume}{58}}, \bibinfo{pages}{940--941}
  (\bibinfo{year}{1991}).

\bibitem{Sque2006}
\bibinfo{author}{Sque, S.~J.}, \bibinfo{author}{Jones, R.} \&
  \bibinfo{author}{Briddon, P.~R.}
\newblock \bibinfo{title}{{Structure, electronics, and interaction of hydrogen
  and oxygen on diamond surfaces}}.
\newblock \emph{\bibinfo{journal}{Phys. Rev. B}} \textbf{\bibinfo{volume}{73}},
  \bibinfo{pages}{085313} (\bibinfo{year}{2006}).

\bibitem{Toyli2010}
\bibinfo{author}{Toyli, D.~M.}, \bibinfo{author}{Weis, C.~D.},
  \bibinfo{author}{Fuchs, G.~D.}, \bibinfo{author}{Schenkel, T.} \&
  \bibinfo{author}{Awschalom, D.~D.}
\newblock \bibinfo{title}{{Chip-scale nanofabrication of single spins and spin
  arrays in diamond}}.
\newblock \emph{\bibinfo{journal}{Nano Lett.}} \textbf{\bibinfo{volume}{10}},
  \bibinfo{pages}{3168--3172} (\bibinfo{year}{2010}).

\bibitem{FavaroDeOliveira2015}
\bibinfo{author}{{F{\'{a}}varo De Oliveira}, F.} \emph{et~al.}
\newblock \bibinfo{title}{{Effect of low-damage inductively coupled plasma on
  shallow nitrogen-vacancy centers in diamond}}.
\newblock \emph{\bibinfo{journal}{Appl. Phys. Lett.}}
  \textbf{\bibinfo{volume}{107}}, \bibinfo{pages}{073107}
  (\bibinfo{year}{2015}).

\bibitem{Pezzagna2010}
\bibinfo{author}{Pezzagna, S.}, \bibinfo{author}{Naydenov, B.},
  \bibinfo{author}{Jelezko, F.}, \bibinfo{author}{Wrachtrup, J.} \&
  \bibinfo{author}{Meijer, J.}
\newblock \bibinfo{title}{{Creation efficiency of nitrogen-vacancy centres in
  diamond}}.
\newblock \emph{\bibinfo{journal}{New J. Phys.}} \textbf{\bibinfo{volume}{12}},
  \bibinfo{pages}{065017} (\bibinfo{year}{2010}).

\bibitem{Deak2014}
\bibinfo{author}{De{\'{a}}k, P.}, \bibinfo{author}{Aradi, B.},
  \bibinfo{author}{Kaviani, M.}, \bibinfo{author}{Frauenheim, T.} \&
  \bibinfo{author}{Gali, A.}
\newblock \bibinfo{title}{{Formation of NV centers in diamond: A theoretical
  study based on calculated transitions and migration of nitrogen and vacancy
  related defects}}.
\newblock \emph{\bibinfo{journal}{Phys. Rev. B}} \textbf{\bibinfo{volume}{89}},
  \bibinfo{pages}{075203} (\bibinfo{year}{2014}).

\bibitem{Goss2005}
\bibinfo{author}{Goss, J.~P.}, \bibinfo{author}{Briddon, P.~R.},
  \bibinfo{author}{Rayson, M.~J.}, \bibinfo{author}{Sque, S.~J.} \&
  \bibinfo{author}{Jones, R.}
\newblock \bibinfo{title}{{Vacancy-impurity complexes and limitations for
  implantation doping of diamond}}.
\newblock \emph{\bibinfo{journal}{Phys. Rev. B}} \textbf{\bibinfo{volume}{72}},
  \bibinfo{pages}{035214} (\bibinfo{year}{2005}).

\bibitem{Takeuchi2005}
\bibinfo{author}{Takeuchi, D.} \emph{et~al.}
\newblock \bibinfo{title}{{Direct observation of negative electron affinity in
  hydrogen-terminated diamond surfaces}}.
\newblock \emph{\bibinfo{journal}{Appl. Phys. Lett.}}
  \textbf{\bibinfo{volume}{86}}, \bibinfo{pages}{152103}
  (\bibinfo{year}{2005}).

\bibitem{RgenRistein2006}
\bibinfo{author}{Ristein, J.}
\newblock \bibinfo{title}{{Surface science of diamond: Familiar and amazing}}.
\newblock \emph{\bibinfo{journal}{Surf. Sci.}} \textbf{\bibinfo{volume}{600}},
  \bibinfo{pages}{3677--3689} (\bibinfo{year}{2006}).

\bibitem{Koslowski2001}
\bibinfo{author}{Koslowski, B.}, \bibinfo{author}{Strobel, S.} \&
  \bibinfo{author}{Ziemann, P.}
\newblock \bibinfo{title}{{Comment on “origin of surface conductivity in
  diamond”}}.
\newblock \emph{\bibinfo{journal}{Phys. Rev. Lett.}}
  \textbf{\bibinfo{volume}{87}}, \bibinfo{pages}{209705}
  (\bibinfo{year}{2001}).

\bibitem{Grotz2012a}
\bibinfo{author}{Grotz, B.} \emph{et~al.}
\newblock \bibinfo{title}{{Charge state manipulation of qubits in diamond}}.
\newblock \emph{\bibinfo{journal}{Nature Communications}}
  \textbf{\bibinfo{volume}{3}}, \bibinfo{pages}{729} (\bibinfo{year}{2012}).

\bibitem{Newell2016}
\bibinfo{author}{Newell, A.~N.}, \bibinfo{author}{Dowdell, D.~A.} \&
  \bibinfo{author}{Santamore, D.~H.}
\newblock \bibinfo{title}{{Surface effects on nitrogen vacancy centers
  neutralization in diamond}}.
\newblock \emph{\bibinfo{journal}{J. Appl. Phys.}}
  \textbf{\bibinfo{volume}{120}}, \bibinfo{pages}{185104}
  (\bibinfo{year}{2016}).

\bibitem{Sato2013c}
\bibinfo{author}{Sato, H.} \& \bibinfo{author}{Kasu, M.}
\newblock \bibinfo{title}{{Maximum hole concentration for Hydrogen-terminated
  diamond surfaces with various surface orientations obtained by exposure to
  highly concentrated NO2}}.
\newblock \emph{\bibinfo{journal}{Diam. Relat. Mater.}}
  \textbf{\bibinfo{volume}{31}}, \bibinfo{pages}{47--49}
  (\bibinfo{year}{2013}).

\bibitem{Schreyvogel2016}
\bibinfo{author}{Schreyvogel, C.} \emph{et~al.}
\newblock \bibinfo{title}{{Active and fast charge-state switching of single NV
  centres in diamond by in-plane Al-Schottky junctions}}.
\newblock \emph{\bibinfo{journal}{Beilstein Journal of Nanotechnology}}
  \textbf{\bibinfo{volume}{7}}, \bibinfo{pages}{1727--1735}
  (\bibinfo{year}{2016}).

\bibitem{lüth2014solid}
\bibinfo{author}{L{\"u}th, H.}
\newblock \emph{\bibinfo{title}{Solid Surfaces, Interfaces and Thin Films}}.
\newblock Graduate Texts in Physics (\bibinfo{publisher}{Springer International
  Publishing}, \bibinfo{year}{2014}).

\bibitem{StreetmanBanerjee}
\bibinfo{author}{Streetman, B.~G.} \& \bibinfo{author}{Banerjee, S.}
\newblock \emph{\bibinfo{title}{{Solid state electronic devices}}}
  (\bibinfo{publisher}{Upper Saddle River, N.J. : Prentice Hall},
  \bibinfo{year}{2000}), \bibinfo{edition}{5th ed} edn.

\bibitem{Fukushima2015a}
\bibinfo{author}{Fukushima, T.}
\newblock \bibinfo{title}{{Precise and fast computation of generalized
  Fermi-Dirac integral by parameter polynomial approximation}}.
\newblock \emph{\bibinfo{journal}{Appl. Math. Comput.}}
  \textbf{\bibinfo{volume}{270}}, \bibinfo{pages}{802--807}
  (\bibinfo{year}{2015}).

\bibitem{Fukushima2015}
\bibinfo{author}{Fukushima, T.}
\newblock \bibinfo{title}{{Precise and fast computation of Fermi-Dirac integral
  of integer and half integer order by piecewise minimax rational
  approximation}}.
\newblock \emph{\bibinfo{journal}{Appl. Math. Comput.}}
  \textbf{\bibinfo{volume}{259}}, \bibinfo{pages}{708--729}
  (\bibinfo{year}{2015}).

\bibitem{Maddox2017}
\bibinfo{author}{Maddox, S.~J.}
\newblock \bibinfo{title}{A free, open-source python package for quickly and
  precisely approximating fermi-dirac integrals.}
\newblock \emph{\bibinfo{journal}{https://pypi.org/project/fdint/}}
  (\bibinfo{year}{2017}).

\bibitem{Pham2015}
\bibinfo{author}{Pham, L.~M.} \emph{et~al.}
\newblock \bibinfo{title}{{NMR Technique for Determining the Depth of Shallow
  Nitrogen-Vacancy Centers in Diamond}}.
\newblock \emph{\bibinfo{journal}{Physical Review B}}
  \textbf{\bibinfo{volume}{93}}, \bibinfo{pages}{045425}
  (\bibinfo{year}{2015}).

\end{thebibliography}

\end{document}